\documentclass[aip,jcp,reprint]{revtex4-1}

\usepackage{graphicx}
\usepackage{subfigure}
\usepackage{amsmath, amsthm}

\begin{document}

\title{Switch-like surface binding of competing multivalent particles}
\author{Nicholas B. Tito}
	\affiliation{Department of Chemistry, University of Cambridge, Lensfield Road, Cambridge CB2 1EW, UK}
	\email{nicholas.b.tito@gmail.com}
\author{Daan Frenkel}
	\affiliation{Department of Chemistry, University of Cambridge, Lensfield Road, Cambridge CB2 1EW, UK}

\begin{abstract}
Multivalent particles competing for binding on the same surface can exhibit switch-like behaviour, depending on the concentration of receptors on the surface. When the receptor concentration is low, energy dominates the free energy of binding, and particles having a small number of strongly-binding ligands preferentially bind to the surface. At higher receptor concentrations, multivalent effects become significant, and entropy dominates the binding free energy; particles having many weakly-binding ligands preferentially bind to the surface. Between these two regimes there is a ``switch-point'', at which the surface binds the two species of particles equally strongly. We demonstrate that a simple theory can account for this switch-like behaviour and present numerical calculations that support the theoretical predictions. We argue that binding selectivity based on receptor density, rather than identity, may have practical applications. 
\end{abstract}

\maketitle

\section{Introduction}

Systems comprising many different chemical ingredients may exhibit complex physical behaviour that is absent in systems with fewer components. In fact, in many biological or bio-mimetic systems, variations in chemical composition allow the system to switch from one mode of operation to another. To give a specific example:  living cells are sensitive to the nature of the molecules bound to their surface, but what molecules bind to a cell surface depends on the chemical composition of that surface~\cite{Mammen:1998im}. A non-biological example is the self-assembly of an elaborate structure from many unique pieces of deoxyribonucleic acid (DNA) \cite{Ke:2012jd, Halverson:2013ge, Reinhardt:2014dt, Hedges:2014iy, Jacobs:2015gi}: here the chemical composition of the system determines the shape of the units that form through self-assembly.

``Multivalent'' particles are a broad class of supra-molecular structures that interact through multiple ligands with (multiple) chemically complementary targets (``receptors''). While the individual ligand-receptor bonds are typically weak, the multiplicity of possible host-guest binding combinations causes the overall binding strength of the particle to be extremely sensitive to system conditions, such as temperature, pH, the number of ligands on the particle, and the receptor concentration on the targeted surface. The binding is \emph{superselective} when the number of adsorbed particles increases faster than linearly with the concentration of receptors \cite{MartinezVeracoechea:2011kn}. 

For a monovalent particle, binding is controlled by the strength of only one ligand-receptor bond. In that case, the number of adsorbed particles depends at most linearly  on the surface density of available receptors. In contrast, for multivalent particles, the 
number of adsorbed particles may depend faster than linear  on the surface density of available receptors. However, provided that the binding strength per ligand-receptor pair is small enough, the multivalent binding to a surface can still be reversible. 

The physical origin of the high sensitivity of multivalent binding to the receptor density is entropic. Typically, unless all bonds are saturated (which only happens for strong binding) there are many ways in which a subset of all ligands can bind to a subset of all receptors. The (logarithm of) the number of ways in which these bonds can be made is an entropic factor that, typically, depends faster than linearly on the receptor concentration. 

Biological systems often exploit multivalency to make bonding of a particle to a target surface extremely sensitive to changes in local environment \cite{Mammen:1998im}. The same holds for man-made multivalent interactions \cite{Varner:2015dh}. 

DNA-coated colloids are an example of synthetic, multivalent building blocks \cite{Mirkin:1996em, Biancaniello:2005ie, Geerts:2010iv, Varilly:2012gl, Michele:2013bw} that can self-assemble. In these systems, complementary single-stranded DNA is grafted to the surfaces of colloids or nanoparticles, resulting in a system that self-assembles over a narrow temperature range into aggregate structures ``encoded'' by the DNA ligands.

Because of their sensitivity to the nature and surface-concentration of receptors. multivalent particles are well suited for 
chemical and biological sensing \cite{Mahon:2014dn}. In some cases, the sensing works by selective break-up of clusters. For instance, gold nanoparticles decorated with bio-responsive ligands have been used as sensing agents to detect the presence of complementary enzymes~\cite{delaRica:2011bt}. In the absence of these enzymes, the nano-particles form clusters via multivalent interactions between their ligands; however, this interaction is disrupted by the presence of the enzyme. As a result, the size of the nano-particle clusters (and hence their plasmon resonance)  depends sensitively on the enzyme concentration.

In the present paper we show that,  depending on their surface concentration, one and the same type of receptor can induce the binding of different multivalent particles. Varying the surface concentration of receptors acts as a ``switch'' between binding of two (or, at least in principle, more) types of multivalent particles. 

To illustrate how such a switch works, we consider a flat surface covered with varying amounts of receptor molecules,  in contact with a reservoir containing two distinct multivalent species. The multivalent particles are assumed  to be spherical and coated with mobile flexible ligands; however, qualitatively,  our conclusions depend neither on the shape of the particles, nor on the mobility of the ligands.

The outline of the remainder of this paper is as follows. We first develop a simple theory to predict the competitive adsorption of two multivalent species by estimating the equilibrium free energy of binding. Following that, we more accurately estimate the free energy of binding with a lattice model, accounting for the change in conformational entropy of a ligand when it binds to a receptor. In the third section we discuss the results obtained from the lattice model, illustrating the influence of energy and entropy on multivalent particle binding, and comparing our results with the relevant theoretical predictions. The paper concludes by discussing possible applications of multivalent switches.

\section{Theory}

The competition for surface adsorption between two multivalent species can be illustrated by considering a simple Langmuir adsorption model \cite{MartinezVeracoechea:2011kn}. In order to develop a mathematically tractable model, we make the following assumptions: the two multivalent particles have approximately equal sizes, and equal ligand lengths; the number of receptors available for binding when a particle is adjacent to the surface is much larger than the number of ligands on the particle; and the binding strength of both particle species is moderate to strong. We will relax these assumptions when examining detailed lattice model calculations in subsequent sections.

Consider a system containing a surface covered by randomly-placed receptors. A reservoir of two multivalent species is in contact with the surface; the two species are identified by $i$ and $j$. A particle of species $i$ has $N_{L,i}$ ligands, and the free energy of binding a single ligand to a receptor is $\Delta f_i$. Species $j$ similarly has parameters $N_{L,j}$ and $\Delta f_j$. When a particle belonging to either of the two species is adjacent to the surface, its ligands are able to access $N_R$ receptors in its vicinity.

The strength of the surface binding of either species depends on the value of $N_{L}$, $\Delta f$, $N_R$ and on the   temperature $T$.
All other things being equal,  the species that has the lower free energy of surface binding will preferentially bind to the surface. Thus, by changing $(N_L, \Delta f)$ for the two species, we may introduce a competition for surface occupancy, and fine-tune which species binds for a given choice of global conditions $(T, N_R)$.

Following ref.~\onlinecite{MartinezVeracoechea:2011kn}, we can estimate the multivalent binding free energy as
\begin{align}
	\nonumber
	\beta F_b &\approx N_L \left(\beta \Delta f - \ln{N_R}\right) \\
	\label{eqn:ValenceScalingPrediction}
	&= - N_L \ln{N_R} + \mbox{const},
\end{align}
where $\beta = 1 / k_B T$ and $k_B$ is the Boltzmann constant. In Eq. \ref{eqn:ValenceScalingPrediction}, we see that the free energy is proportional to $\ln{N_R}$, with a scaling factor of $N_L$. The binding free energies of species $i$ and $j$ are equal, i.e. $F_{b, i} = F_{b, j}$, when,
\begin{equation}
	\left( k_B T \ln{N_R} \right)_{\mbox{switch}} = \frac{N_{L,j} \Delta f_j - N_{L,i} \Delta f_i}{N_{L,j} - N_{L,i}}.
	\label{eqn:SwitchPoint}
\end{equation}
The relation given by Eq. \ref{eqn:SwitchPoint} defines a unique choice of the global parameters $(T, N_R)$ at which the two multivalent species bind equally strongly to the surface. In what follows, we will assume that the is temperature fixed: we will only  vary the receptor count $N_R$. 

Let us now consider the case that species $i$ has a small number of strongly binding ligands, while species $j$ has many less-strongly-binding ligands;
\begin{align}
	\nonumber
	& N_{L,j} \gg N_{L,i} \\
	\nonumber
	& \Delta f_i \ll \Delta f_j.
\end{align}
In this regime, we can still use Eq. \ref{eqn:ValenceScalingPrediction} to predict how the difference in binding free energies between the two species depends on $N_R$:
\begin{equation}
	\nonumber
	\beta \Delta F_{ji,b} \approx \beta \left(N_{L,j} \Delta f_j - N_{L,i} \Delta f_i\right) - N_{L,j} \ln{N_R}.
\end{equation}
We note that the binding  free energy of species $i$, with a small number of strongly binding ligands, has a weak dependence on $N_R$ in this limit. However, the binding free energy of species $j$, with many weaker binding ligands, depends strongly on $\ln{N_R}$.  When $N_R$ is large, then $F_{b,j} > F_{b,i}$, as the \emph{entropy} of binding is the dominant factor in the bound-state free energy of the two species.   On the other hand, when $N_R$ is small, then $F_{b,i} > F_{b,j}$, and the \emph{energy} of binding determines which species binds to the surface. When $N_R$ satisfies Eq. \ref{eqn:SwitchPoint}, then the entropy and energy of binding of each species are at a balance.

The simple theory here reveals that switch-like behaviour occurs when the binding free energy for each species depends differently on the surface-receptor concentration. Therefore, mixtures of monovalent species will never exhibit surface switching. For example, if a system contains two monovalent species, one strongly binding, and the other weakly binding, then the ratio of the surface concentrations of the two species will not depend on the receptor concentration.

\section{Lattice Model for Ligand-Receptor Binding}

\begin{figure}
	\centering
		\subfigure[]{\includegraphics[width=0.495\textwidth]{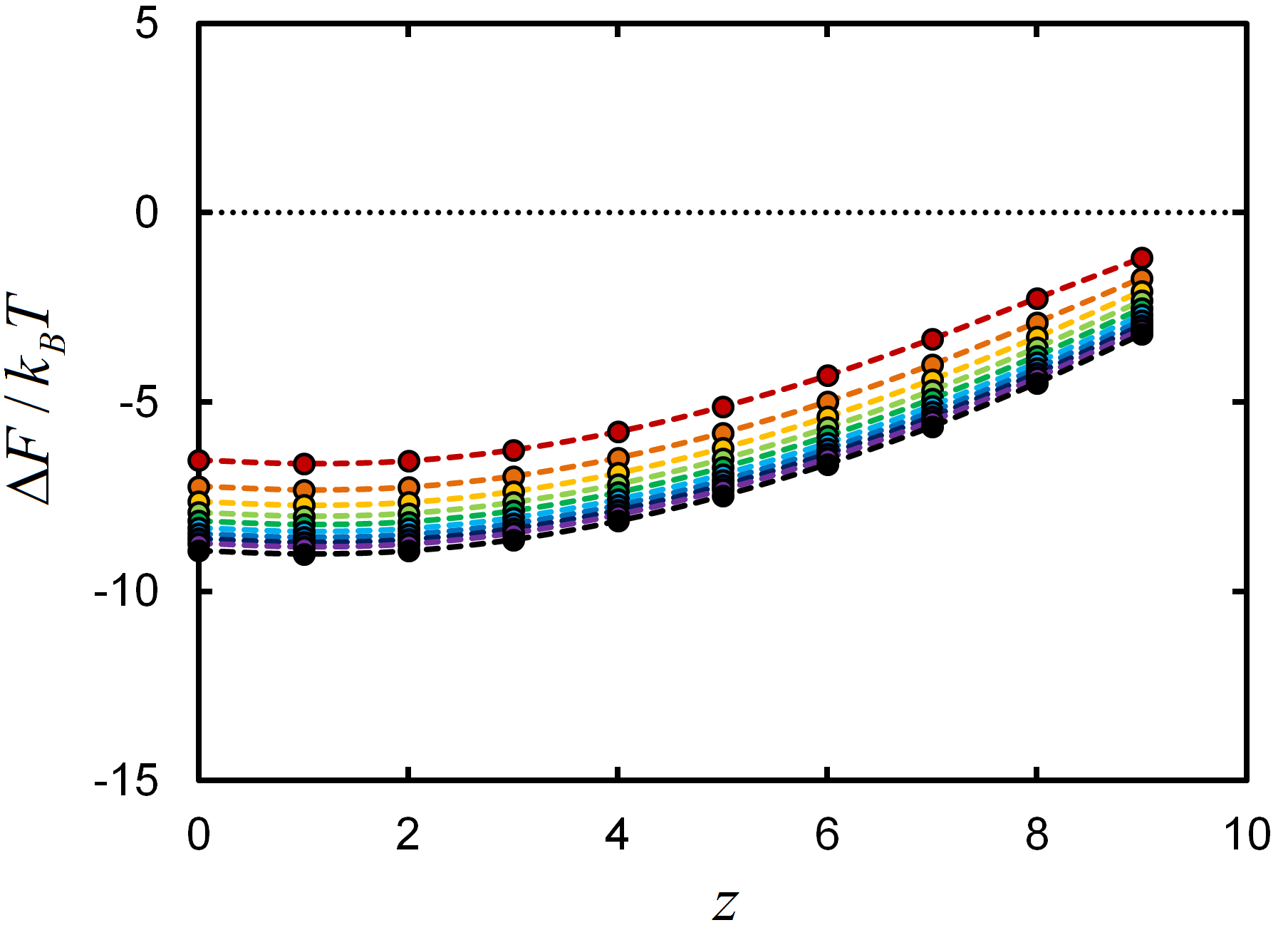}}
		\subfigure[]{\includegraphics[width=0.495\textwidth]{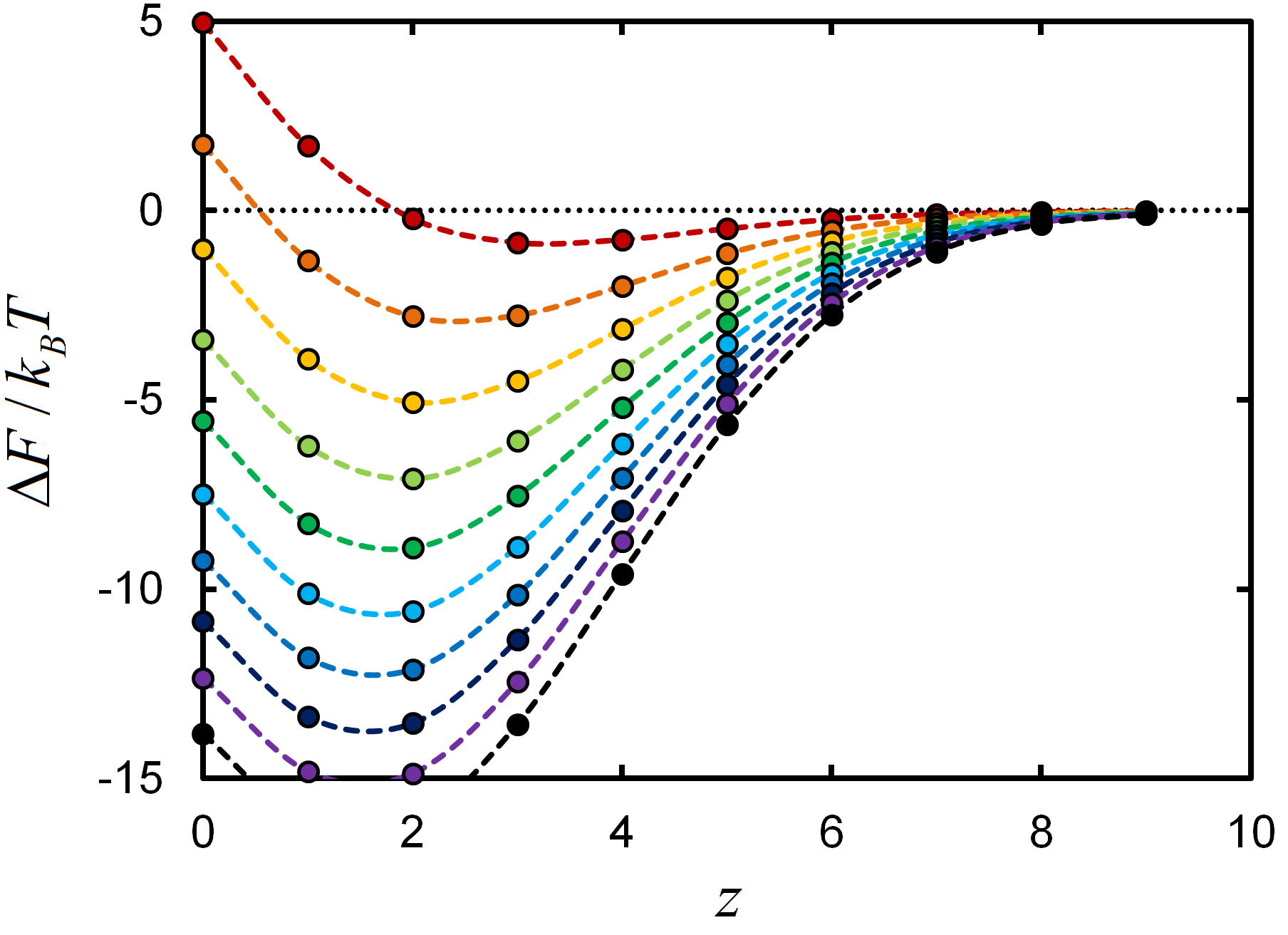}}
		\caption{Binding free energy $\Delta F /  k_B T$ as a function of distance $z$ between receptor and particle surfaces, for different choices of receptor concentration $\phi_R$. Values of receptor concentration increase from red to dark violet, progressing 0.1 to 1.0 in increments of 0.1. Points are numerical results, and lines are guides to the eye. In (a), $N_{L} = 1$ and $\epsilon = -12 k_B T$; in (b), $N_{L} = 20$ and $\epsilon = -3.5 k_B T$. Both particles have ligands of $N_{poly} = 20$ segments, and cores of radius $r = 2$.}
	\label{fig:FEvsH-Shifted}
\end{figure}

In the previous section, a key parameter was $\Delta f$, the binding free energy per  ligand. This free energy has two contributions:  the binding energy of the ligand-receptor interaction and the difference in configurational entropy of the ligand in the bound and unbound states.  This configurational entropy depends on the nature of the ligand and on the space accessible to the ligand in its bound and unbound states. 

To compute the approximate entropy change of a ligand upon binding, we use a lattice model. The model represents a single multivalent particle with ligands, interacting with receptors placed on a surface. We account for the fact that the ligands cannot penetrate the surface and that no two ligands can be bound to the same receptor. In ref.~\onlinecite{CompanionPaperRef} we describe an efficient method to compute the binding free energy of a multivalent particle, taking into account the above constraints.

We represent the system on a three-dimensional simple cubic lattice. A flat impenetrable surface is located at vertical coordinate $h = 1$ (in lattice units), on which receptor sites are randomly placed. The core of the multivalent particle is represented on the lattice by a (discretised) hard sphere, with radius $r$, and center located at $(x^*, y^*, h^*)$. Impenetrability is enforced by preventing any ligand segments from occupying lattice sites with coordinates $(x, y, h)$ satisfying $\left(x - x^*\right)^2 + \left(y - y^*\right)^2 + \left(h - h^*\right)^2 \leq r^2$. Given that the receptor surface is located at $h = 1$, the distance between the receptor surface and particle surface is $z = h^* - r - 2$.

The ligands of the multivalent particle are represented as non-self-avoiding lattice walks of $N_{poly}$ steps. Recursive enumeration \cite{Rubin1965} is used to calculate the number $q'_j$ of walks that extend from anywhere on the particle core surface, to site $j$ on the receptor surface. The energy of a single ligand-receptor bond is $\epsilon$ for all receptors; the total binding weight for site $j$ is therefore $q_j = q'_j \exp{(-\beta \epsilon)}$. Each of the $N_{L}$ ligands belonging to the particle are assumed to be chemically identical, so that there is no dependence on ligand identity when computing the binding weights $q_j$. The partition function $q_{ub}$ per unbound ligand is defined to be the number of non-self-avoiding walks beginning on the particle surface and ending anywhere in the system. Each $q_j$ as well as $q_{ub}$ depend on the spacing $z$ between the particle and receptor surfaces.

We denote by $N_A(z)$ the number of surface sites accessible to the particle ligands, when the particle has spacing $z$ from the receptors surface. $N_A(z)$ is determined by geometry: for a particle having ligands of length $N_{poly}$, at a distance $z$ from the surface $N_A(z) \approx 2 (N_{poly} - z + 1)^2$ and the mean number of receptors available to the particle is
\begin{equation}
	\nonumber
	\bar{N}_R(z) = N_A(z) \phi_R,
\end{equation}
where $\phi_R$ is the probability to find a receptor at a surface site.  

To calculate the binding free energy, we must average over all possible surface receptor configurations. We assume that receptors are randomly distributed over the $N_A(z)$ accessible surface sites. The probability $P(N_R)$ to find exactly $N_R$ receptors on $N_A(z)$ sites is then:
\begin{equation}
	\nonumber
	P(N_R) = \binom{N_A}{N_R} \phi_R^{N_R} \left(1 - \phi_R\right)^{N_A - N_R},
\end{equation}
where we have written $N_A$ for $N_A(z)$. From ref.~\onlinecite{CompanionPaperRef}, the bound-state partition function for $z$ averaged over every possible receptor configuration is
\begin{equation}
	Q_{full} = \sum_{\lambda = 0}^{\min{(N_A,N_L)}} Q(\lambda),
\end{equation}
where
\begin{align}
	Q(\lambda) = &\frac{\binom{N_L}{\lambda}}{\binom{N_A}{\lambda}} \lambda! Q_{ub}(\lambda) Q_b(N_A, \lambda) \times \sum_{N_R = \lambda}^{N_A}{\binom{N_R}{\lambda} P(N_R)}.
	\label{eqn:QBound}
\end{align}
The quantity $Q_b(N_A, \lambda)$ is computed by finding the residue of a complex integral \cite{CompanionPaperRef}. The quantity $Q_{ub}(\lambda) = q_{ub}^{\left(N_L - \lambda\right)}$ is the partition function for the unbound ligands. The bound free energy is then obtained by
\begin{equation}
	F(z) = -k_B T \ln{Q_{full}}.
	\nonumber
\end{equation}
It is useful to calculate this quantity relative to the free energy $F^\circ$ when the particle is at infinite distance from the surface, where $F^\circ = -N_L \ln{q^\circ}$ and $q^\circ$ is the partition function for one ligand in the reference state. The free energy of binding is therefore
\begin{equation}
	\Delta F(z) = F(z) - F^\circ.
	\label{eqn:FEBound}
\end{equation}
Equation \ref{eqn:FEBound} is calculated for different values of $z$. The position $z^*$ that leads to the most negative binding free energy $\Delta F^*$ is taken to be the equilibrium position for fixed $(N_{poly}, N_L, \beta\epsilon, \phi_R)$.

\begin{figure}
	\centering
	\includegraphics[width= 0.50 \textwidth]{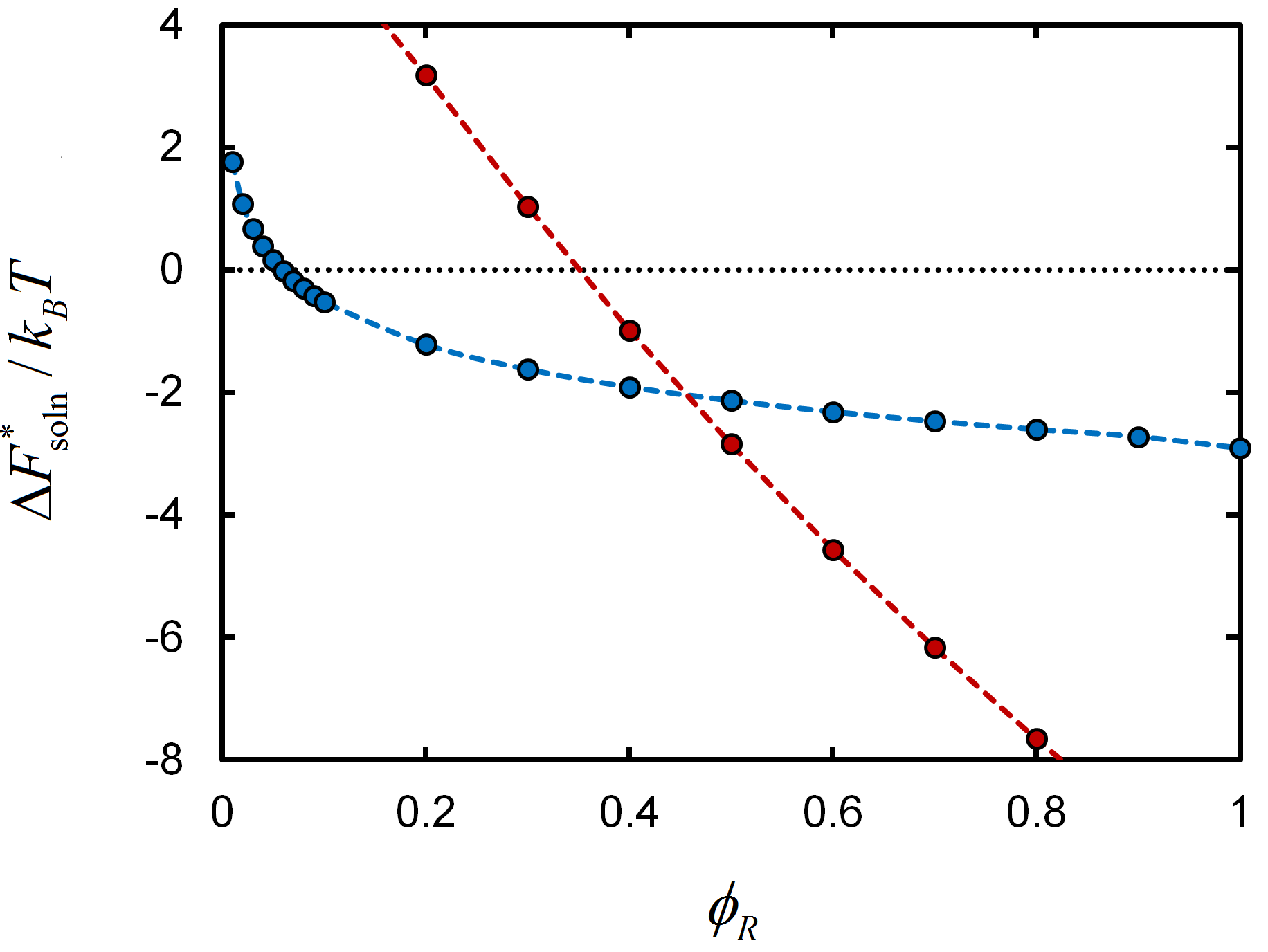}
	\caption{Solution binding free energy $\Delta F^*_{soln} /  k_B T$ vs. surface receptor concentration $\phi_R$ for the two particle species $i$ (blue) and $j$ (red). Solution chemical potential is $\beta \mu_{id} = -6.1$. Points are numerical calculations, and lines are guides to the eye.}
	\label{fig:FEMINvsRCONC}
\end{figure}

\begin{figure}
	\centering
	\includegraphics[width= 0.50 \textwidth]{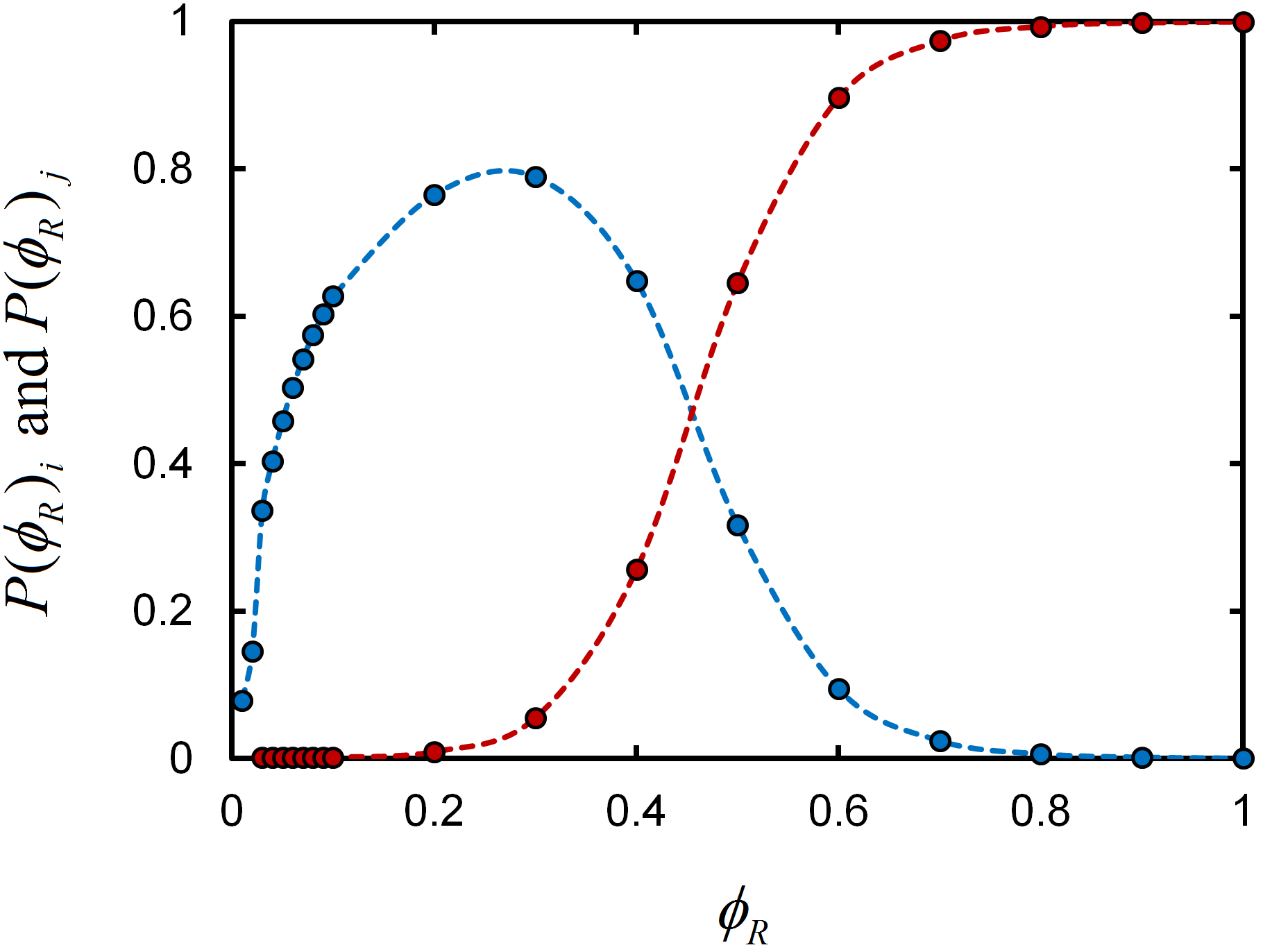}
	\caption{Probability that species $i$ (blue) and $j$ (red) are bound to the surface as a function of receptor concentration. Points are numerical results, and lines are guides to the eye.}
	\label{fig:PBOUNDvsRCONC}
\end{figure}

\section{Results \& Discussion}

We now numerically calculate the free energy of binding for surface-adsorbing multivalent particles, as a function of the concentration of surface receptors. Competition between two particle species is illustrated by comparing their binding free energy to the same surface. In the Supporting Information, we perform a more detailed analysis of how multivalency affects the entropy and energy of binding, including the limit of strong-binding ligands.

Two species of particles are defined in the system: one with few strong-binding ligands, and the other with many weak-binding ones. The first species, $i$, has $N_{L,i} = 1$ and $\epsilon_i = -12 k_B T$, while the second species $j$ has $N_{L,j} = 20$ and $\epsilon_j = -3.5 k_B T$. While quantitatively, our results depend on the parameters chosen, the broader concept of switch-like competition between the two species does not.

The results in Figure \ref{fig:FEvsH-Shifted} show the free energy of binding as a function of distance $z$ between the particle and receptor surface. Results are shown for different choices of receptor concentration $\phi_R$ at fixed temperature. 

Figure \ref{fig:FEvsH-Shifted}(a) indicates that the low-valence particle with one strong-binding ligand has binding free energy minima very near the receptor surface. The value of the binding free energy at the minimum changes little with receptor concentration. On the other hand, in Figure \ref{fig:FEvsH-Shifted}(b), the particle with many weak-binding ligands has a preferred binding position some distance away  from the surface, due to the effective repulsion caused by unbound ligands. With increasing receptor concentration the minimum in the free energy curve shifts closer to the surface.  Moreover, the minimum in the binding free energy becomes deeper as $\phi_R$ increases.

We now consider the situation where the bulk concentration of each species is $\approx 0.002$ particles per unit volume, leading to an (ideal) chemical potential of $\beta \mu_{id} = -6.1$. The free energy of binding of a particle from this hypothetical solution is calculated by
\begin{equation}
	 \Delta F_{soln}^* = \Delta F^* - \mu_{id},
	 \nonumber
\end{equation}
which incorporates the solution chemical potential, and the minimum free energies of binding from Figure \ref{fig:FEvsH-Shifted}.

In Figure \ref{fig:FEMINvsRCONC}, solution free energies of binding $\Delta F_{soln}^*$ are plotted as a function of $\phi_R$. At low receptor concentration, the low-valence species $i$ has a more negative (more favourable) binding free energy compared to the high-valence species $j$. As the receptor concentration grows larger, the binding free energy of the high-valence species becomes more favourable. More receptors result in more possible ligand-receptor binding permutations for species $j$, causing the entropic contributon to the binding free energy for that species to become more significant. Near a receptor concentration of $0.5$, the two curves cross at a ``switch point'', and at larger $\phi_R$ the high-valence species is more favourably bound to the surface.

Figure \ref{fig:PBOUNDvsRCONC} illustrates the switch point of this system in terms of the probability that a particular species is bound to the surface. For species $i$ for example,
\begin{equation}
	P(\phi_R)_i = \frac{e^{-\beta \Delta F^*_{soln,i}(\phi_R)}}{1 + e^{-\beta \Delta F^*_{soln,i}(\phi_R)} + e^{-\beta \Delta F^*_{soln,j}(\phi_R)}}.
	\nonumber
\end{equation}
where $\Delta F^*_{soln,i}(\phi_R)$ is the solution binding free energy for species $i$ at a receptor concentration of $\phi_R$. At low receptor concentration, species $i$ is bound weakly, consistent with its rather small binding free energy of $ \approx -1 k_B T$. At larger receptor concentration, species $i$ binds more strongly initially, but then is replaced by the more favorably-bound species $j$. 

Multivalent particles interact with receptors on a surface by different ligand-receptor pair combinations. This is like connecting the sockets of an old telephone switchboard in different ways. In the Supporting Information, we isolate and examine the role of ``switchboard entropy''---the entropy associated with different ligand-receptor combinations---in the context of switch-like particle binding competition as seen in Figure \ref{fig:FEMINvsRCONC}.

We next consider how the switch point of two competing multivalent species changes as a function of system parameters, such as valency $N_L$ and binding strength $\epsilon$. Equation \ref{eqn:SwitchPoint} represents a rough prediction for the $N_R$ at which there is coexistence between two species $i$ and $j$. Equation \ref{eqn:SwitchPoint} can be rearranged, separating out the energetic and entropic contributions, to yield
\begin{align}
	\ln{N_{R,\mbox{switch}}} &= \frac{\beta \left(N_{L,j} \epsilon_j - N_{L,i} \epsilon_i\right)}{N_{L,j} - N_{L,i}} \nonumber \\
	& - \frac{N_{L,j} s_j/k_B - N_{L,i} s_i/k_B}{N_{L,j} - N_{L,i}} \nonumber \\
	&= \eta - b.
	\label{eqn:SwitchTheoryFit}
\end{align}
Represented in this way, the theory predicts that the switch-point $\ln{N_R}$ values will be directly proportional to $\eta$ with slope unity, and with an intercept of $b$ representing a binding entropy difference between the two species. The quantity $\eta$ is useful, as it is a function only of the parameters used to define the two multivalent species in the system.

\begin{figure}
	\centering
		\includegraphics[width=0.50 \textwidth]{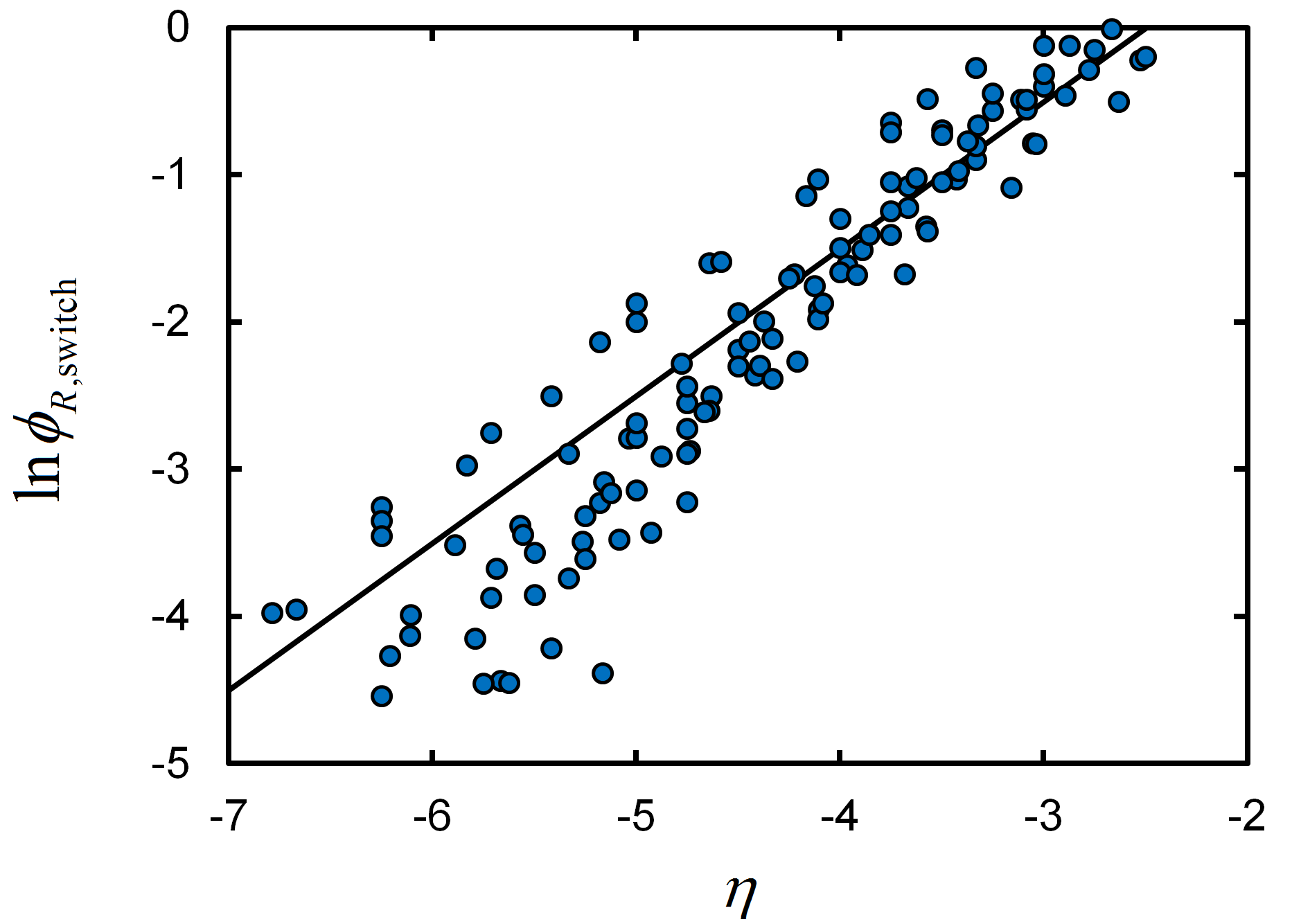}
		\caption{Switchpoint $(\ln{\phi_R})$ as a function of binding parameter $\eta$ for competing multivalent surface adsorbers. Points are numerical calculations for a variety of $\eta$; solid line has slope of $1$, indicating expected scaling between $\eta$ and $\ln{\phi_R}$ given by Eq. \ref{eqn:SwitchTheoryFit}.}
	\label{fig:PHISWvsETA}
\end{figure}

We now show that the switch-point receptor concentration $\phi_R$ scales linearly with the parameter $\eta$ as predicted by Eq. \ref{eqn:SwitchTheoryFit}. Figure \ref{fig:PHISWvsETA} shows values of $(\ln{\phi_R})$ at which surface switching occurs for a variety of pairs of competing multivalent species $i$ and $j$. Choosing $(N_{L, i}, \epsilon_i)$ and $(N_{L, j}, \epsilon_j)$ for species $i$ and $j$ respectively leads to a unique value of the parameter $\eta$, as displayed in Eq. \ref{eqn:SwitchTheoryFit}. 

The results in Figure \ref{fig:PHISWvsETA} suggest that $(\ln \phi_{R,switch})$ scales approximately linearly with $\eta$, as predicted by the simple theory in Eq. \ref{eqn:SwitchTheoryFit}. At larger (more negative) $\eta$, the numerical results deviate from the theory. According to Eq. \ref{eqn:SwitchTheoryFit}, a more negative $\eta$ leads to a smaller value of the switch-point receptor concentration $\phi_R$. Thus, we reasonably expect Eq. \ref{eqn:SwitchTheoryFit} to break down in this limit, given that large $N_R$ was assumed in the derivation in Section 2. Nevertheless, the agreement between the numerical calculations and the theory is surprisingly good, given that the theory also does not account for the polydispersity $P(N_R)$ of the local number of surface receptors.

\section{Conclusions}

The results presented in this paper indicate that the preferred adsorption of particles can be switched by changing surface receptor concentration, provided that one particle has few strong-binding ligands, while the second has many weak-binding ligands. Importantly, this behaviour is unique to multivalent particles, given that a mixture of monovalent species always exhibits the same coexistence ratio on the surface regardless of receptor concentration.

A lattice model coupled with a recent multivalent binding theory \cite{CompanionPaperRef} was employed to develop a more detailed picture of switch-like adsorption. At low receptor concentration, the low-valence species having strong-binding ligands preferentially adsorbs to the surface. At larger receptor concentration, binding entropy associated with multivalency becomes a significant factor in the free energy, causing the high-valence species to occupy the surface instead. In between, there is a unique choice of receptor concentration at which the two species coexist. The coexistence points closely scale with a simple Langmuir theory for the cases where the theory assumptions are valid. 

Examining the energy and entropy of binding in detail (see Supporting Information) reveals the role that receptor degeneracy plays in the binding process. The bound-state entropy of the multivalent species initially \emph{decreases} as the number of receptors increases. This is due to the entropic cost associated with binding ligands to the surface. Once the particle has all of its ligands bound, there is a turning point, and the equilibrium entropy begins to \emph{increase} again due to the availability of excess receptors. 

This ``switchboard'' entropy dominates the free energy of binding of multivalent species when there are excess receptors, while having an insignificant role for monovalent species. Altering the number of ligands on a particle therefore tunes the magnitude of the switchboard entropy term in the binding free energy, and this can be used to drive competitive surface adsorption in a mixture of species.

The multivalent physics discussed in this paper suggest development of a type of ``multivalent chromatography'', in which a sensing surface is used to isolate particles from a polydisperse mixture based on their valency. Recent experimental progress on the development of self-assembled monolayers having chemical gradients is promising \cite{Nicosia:2014bw}, as these would serve as ideal substrates for performing the chromatography. DNA dendrimers \cite{Li:2003fp} present an opportunity to experimentally study surface switching behaviour in a controlled fashion, given the ability to synthesize dendrimers with specific valency and ligand interaction energy. These applications will be the subject of forthcoming work.

\section{Acknowledgments}

The research leading to these results has received funding from the European Research Council under the European Union's Seventh Framework Programme (FP/2007-2013) / ERC Grant Agreement n. 607602 (``SASSYPOL''). Funding is also gratefully acknowledged from EPSRC Programme Grant EP/I001352/1. We wish to thank Tine Curk, Stefano Angioletti-Uberti, and Peter Bolhuis for helpful discussions on this work.

\bibliography{main}

\newpage

\renewcommand\thefigure{S\arabic{figure}}
\renewcommand\thesection{S\Roman{section}}
\renewcommand\theequation{S\arabic{equation}}

\setcounter{figure}{0}
\setcounter{section}{0}
\setcounter{equation}{0}

\begin{widetext}
\noindent \MakeUppercase{\textbf{Supporting Information for ``Switch-like surface binding of competing multivalent particles''}}
\end{widetext}

\section{Entropy and energy of binding}

Here we take a detailed look at how entropy and energy influence multivalent particle binding, giving rise to the overall free energy like observed in Figure 1. The results presented here are calculated using the model in Section 3.

To compute the average energy of binding at a given $z$, we multiply the ensemble average number of bound ligands by the energy $\epsilon$ per bond:
\begin{equation}
	\beta U = \frac{\epsilon}{Q_{full}} \times \sum_{\lambda = 0}^{\min{(N_A, N_L)}}{ \lambda  Q(\lambda) }.
	\nonumber
\end{equation}
The partition functions $Q(\lambda)$ and $Q_{full}$ are given by Eqs. 3 and 4. When the multivalent particle is at infinite distance from the surface, the total energy is $U^\circ = 0$ as there are no ligand-receptor bonds. Therefore, the change in energy upon surface binding is simply $\Delta U = U$. 

\begin{figure*}
	\centering
		\subfigure[]{\includegraphics[width=0.495\textwidth]{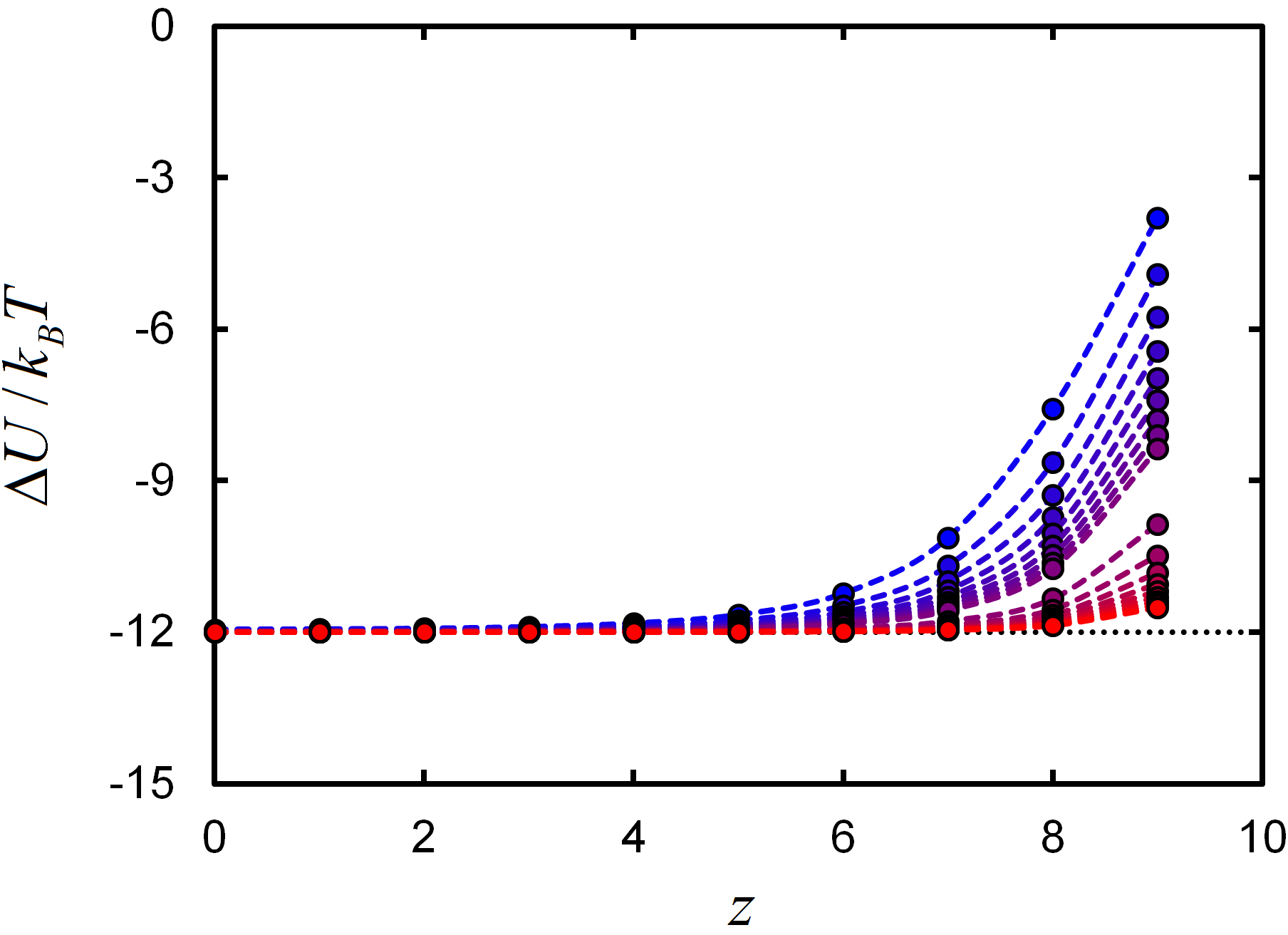}}
		\subfigure[]{\includegraphics[width=0.495\textwidth]{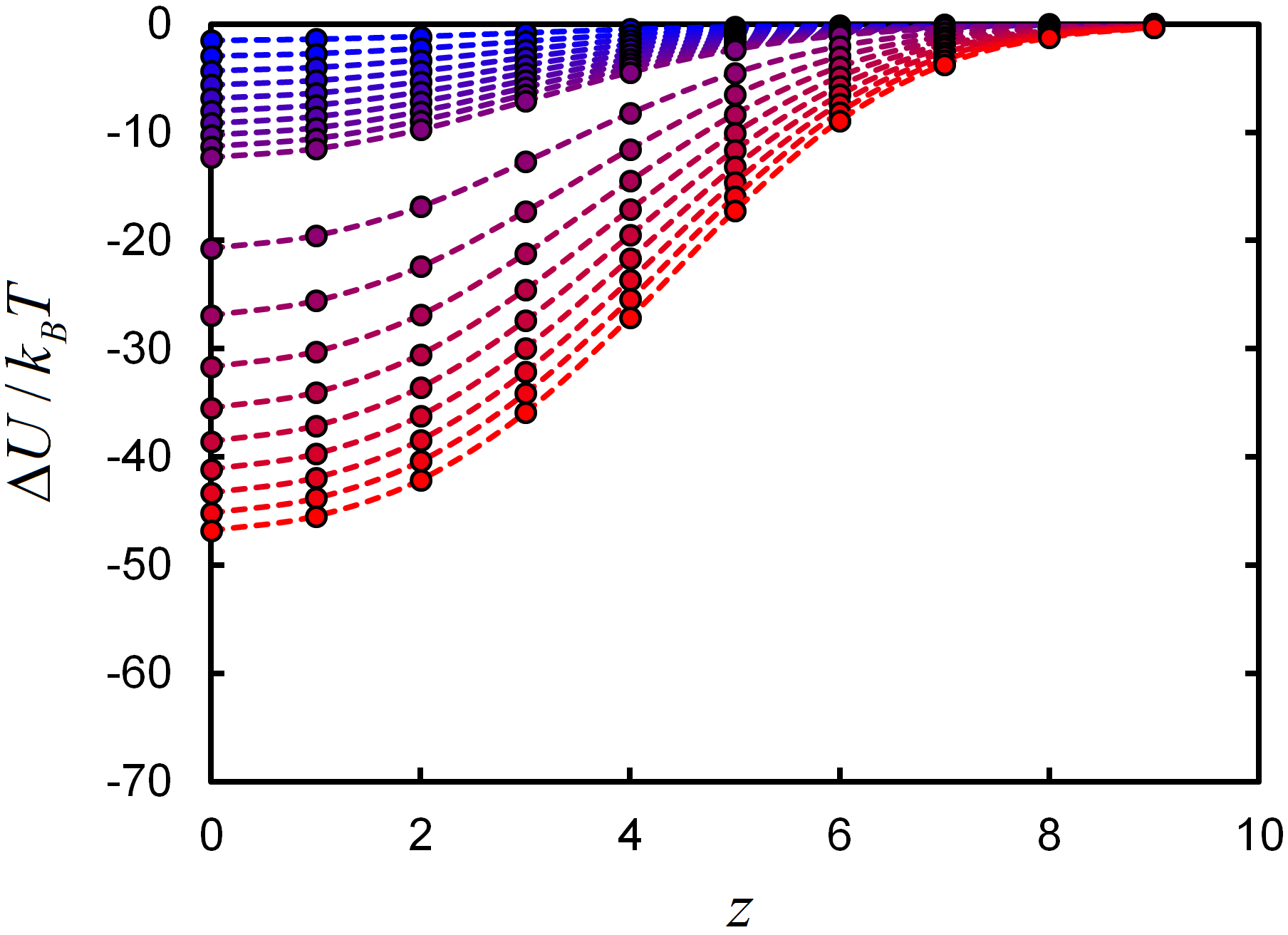}}
		\subfigure[]{\includegraphics[width=0.495\textwidth]{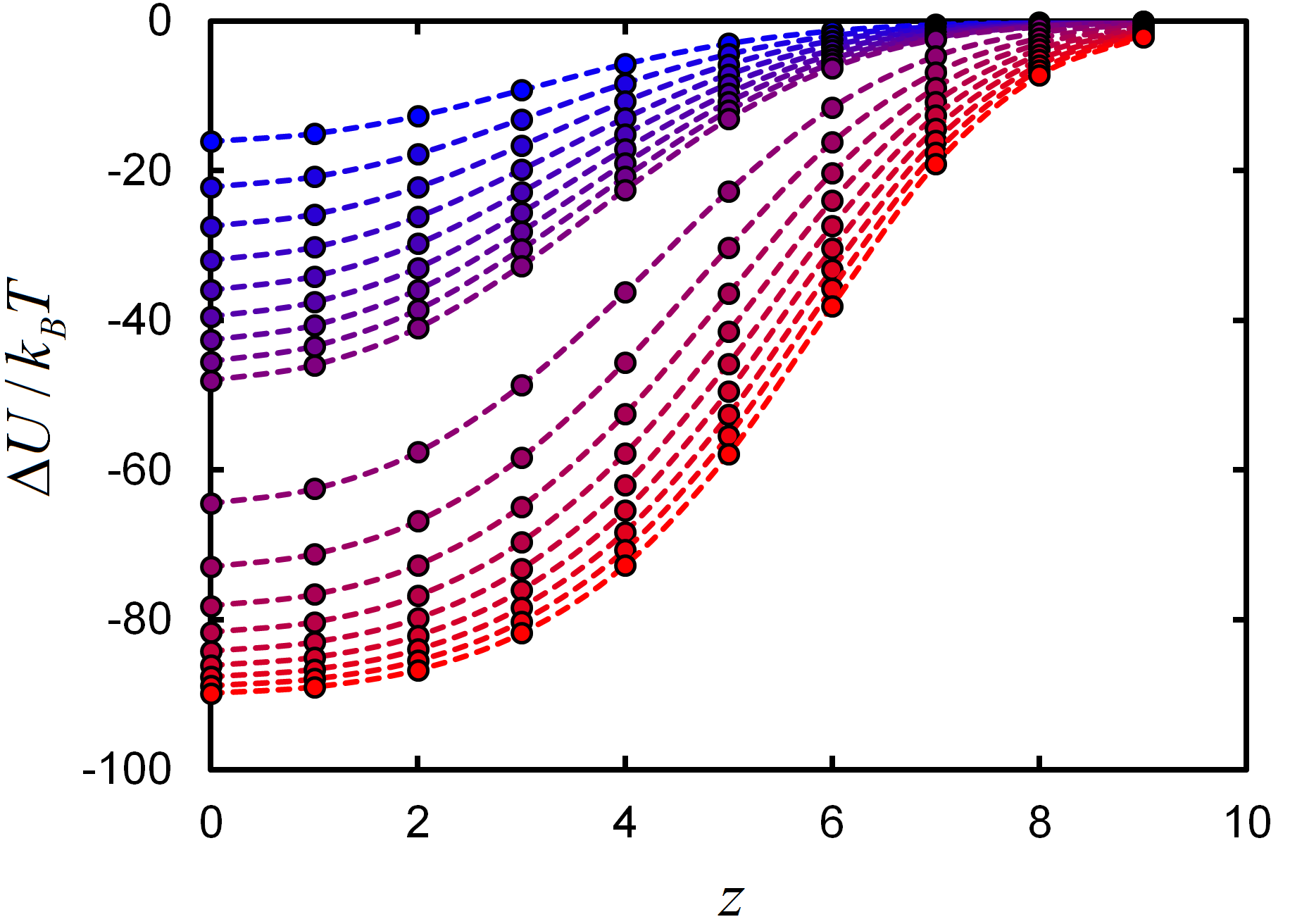}}
		\subfigure[]{\includegraphics[width=0.495\textwidth]{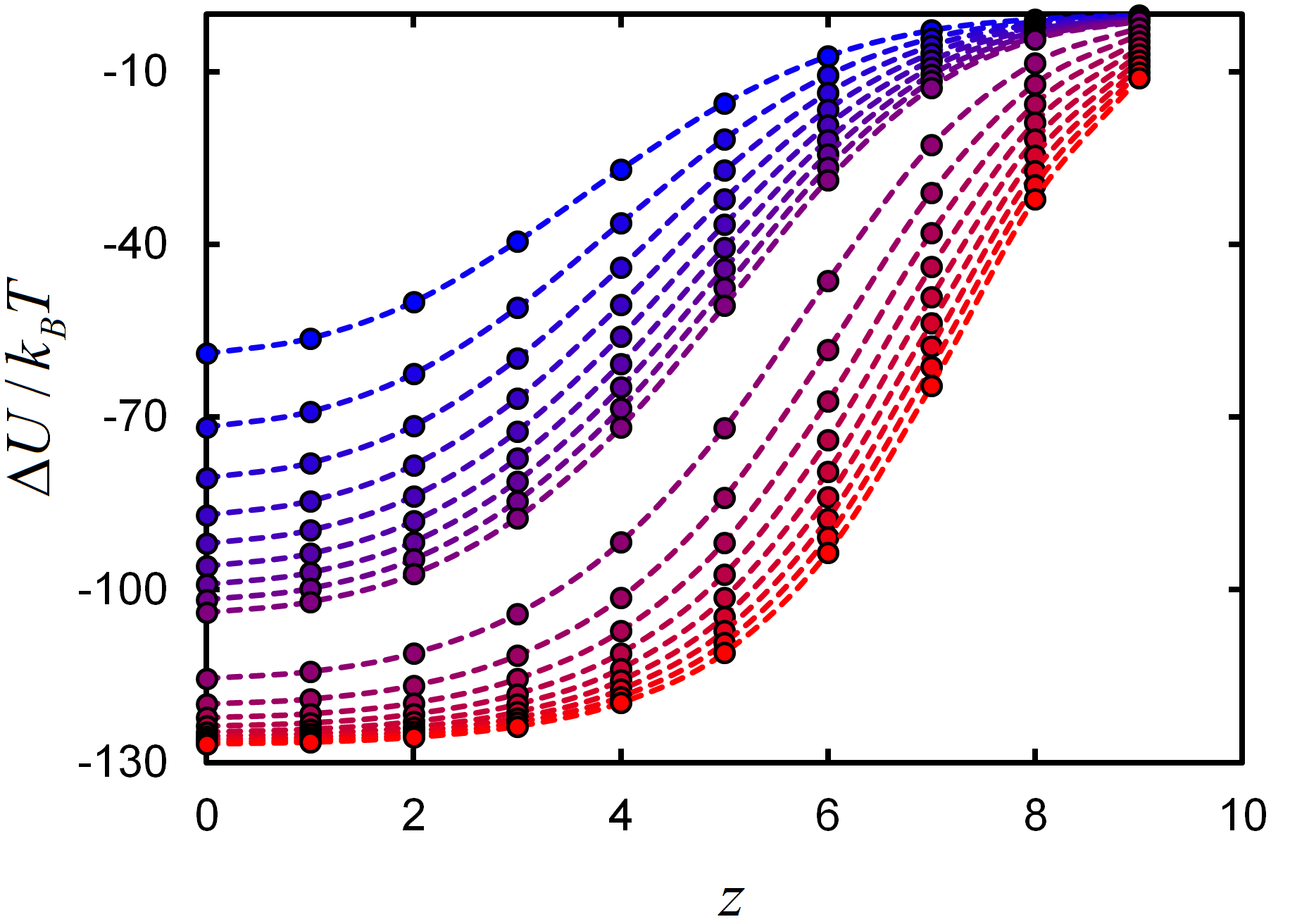}}
		\caption{Energy of binding $\Delta U/k_B T$ as a function of distance $z$ between receptor and particle surfaces, for different choices of receptor concentration $\phi_R$. Values of receptor concentration increase from blue to red, progressing 0.01 to 0.1 in increments of 0.01, and then 0.1 to 1.0 in increments of 0.1. Panel (a) has $N_L = 1$ and $\beta \epsilon = -12$. The remaining panels have $N_L = 20$; values of $\beta \epsilon$ are (b) -3.5, (c) -5, (d) -6.5. Points are numerical results, and lines are guides to the eye.}
	\label{fig:UvsH}
\end{figure*}

Figure \ref{fig:UvsH} gives the binding energy $\beta \Delta U$ as a function of $z$. The first panel shows results for a strong-binding particle with $N_L = 1$, and the remaining are for a particle with $N_L = 20$ ligands and three different ligand-receptor binding energies.

\begin{figure*}
	\centering
		\subfigure[]{\includegraphics[width=0.495\textwidth]{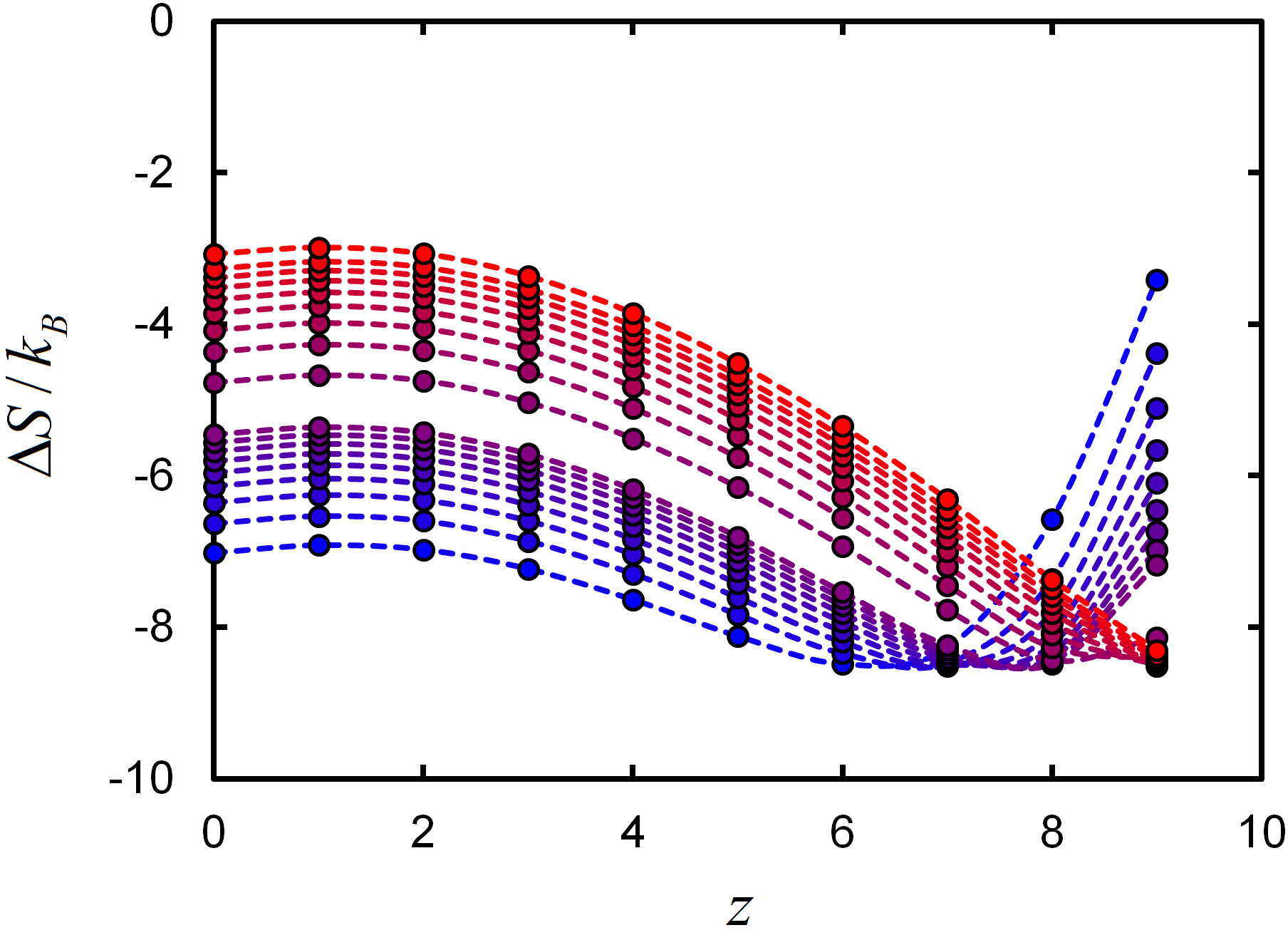}}
		\subfigure[]{\includegraphics[width=0.495\textwidth]{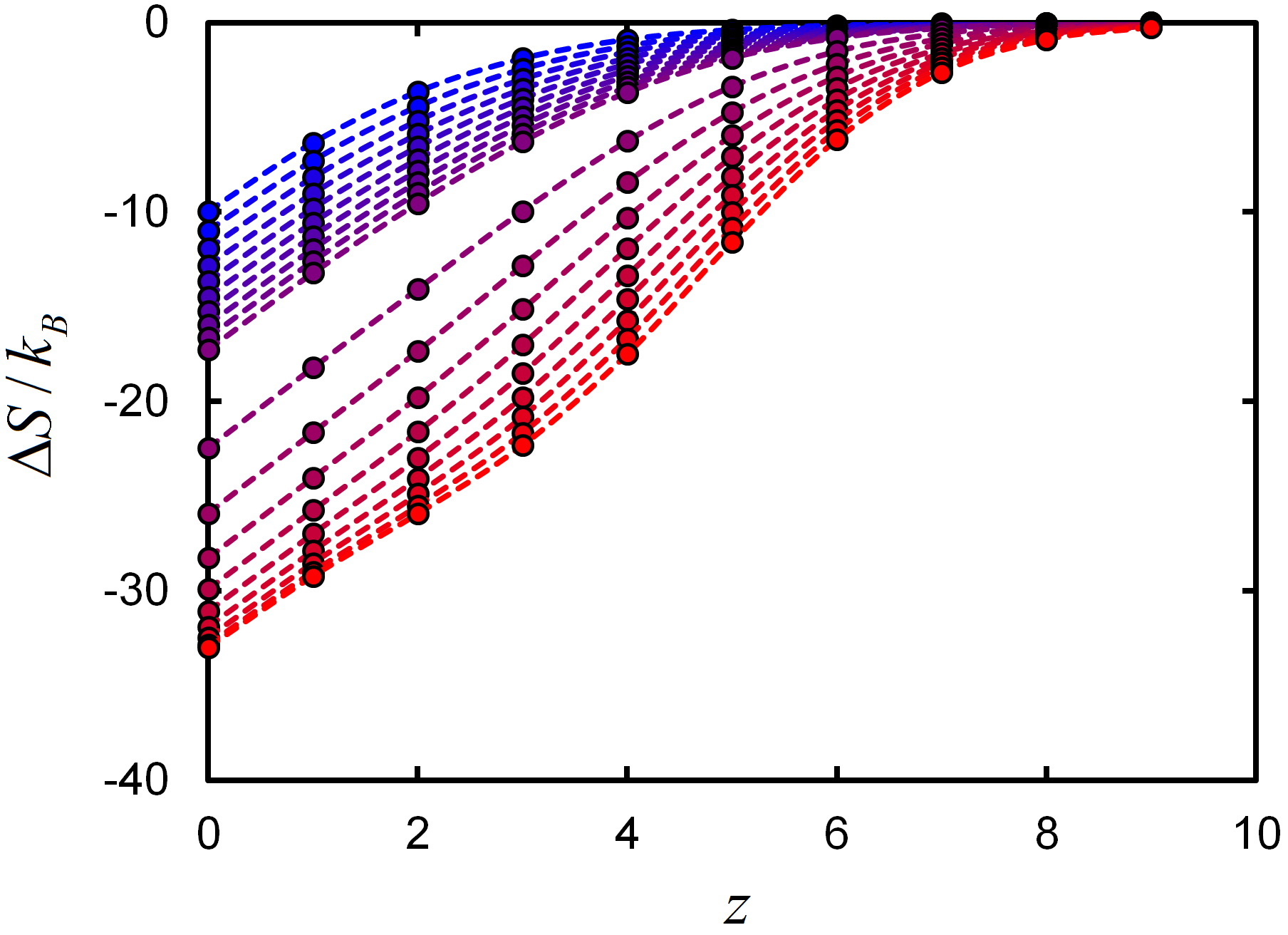}}
		\subfigure[]{\includegraphics[width=0.495\textwidth]{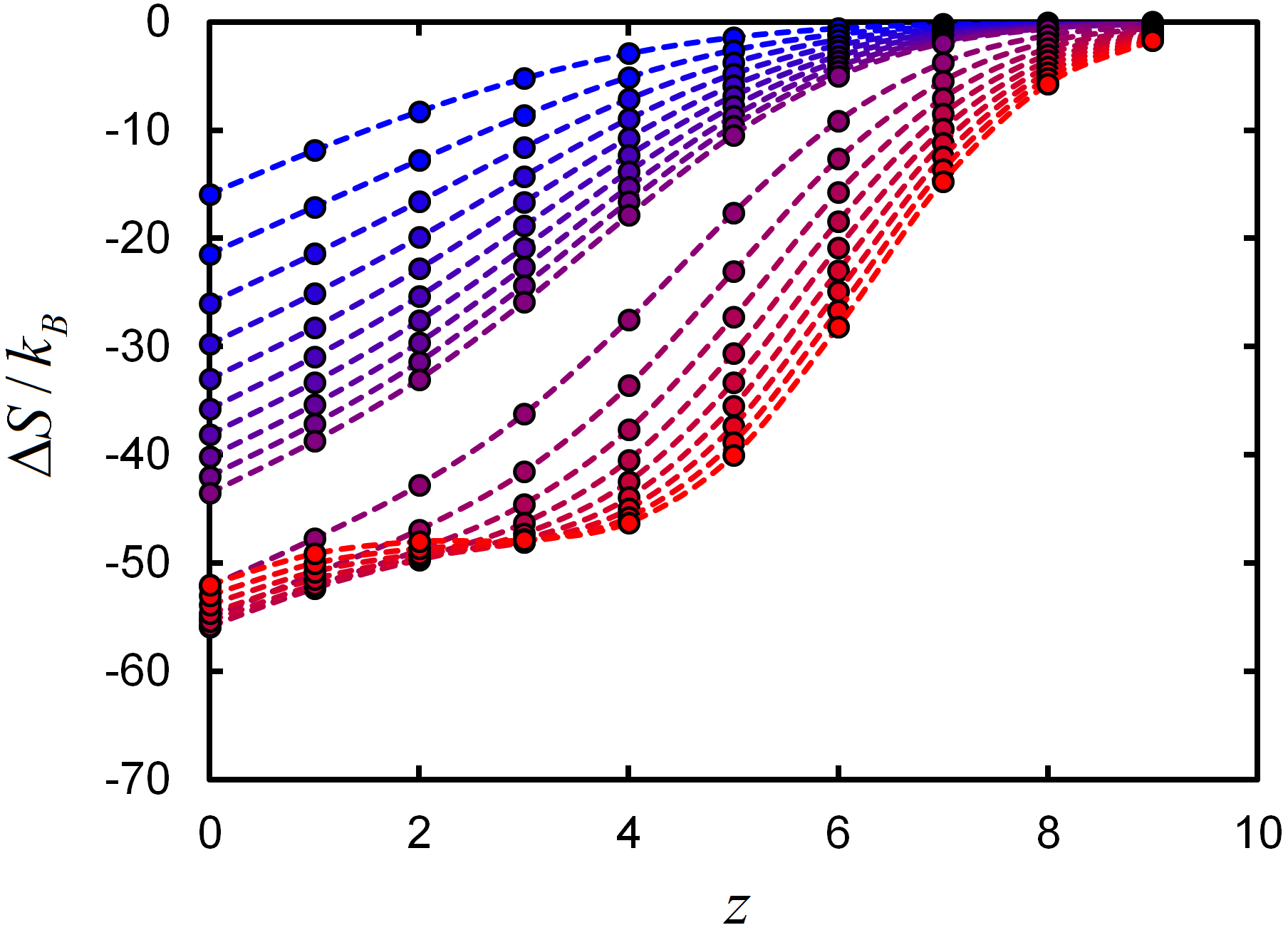}}
		\subfigure[]{\includegraphics[width=0.495\textwidth]{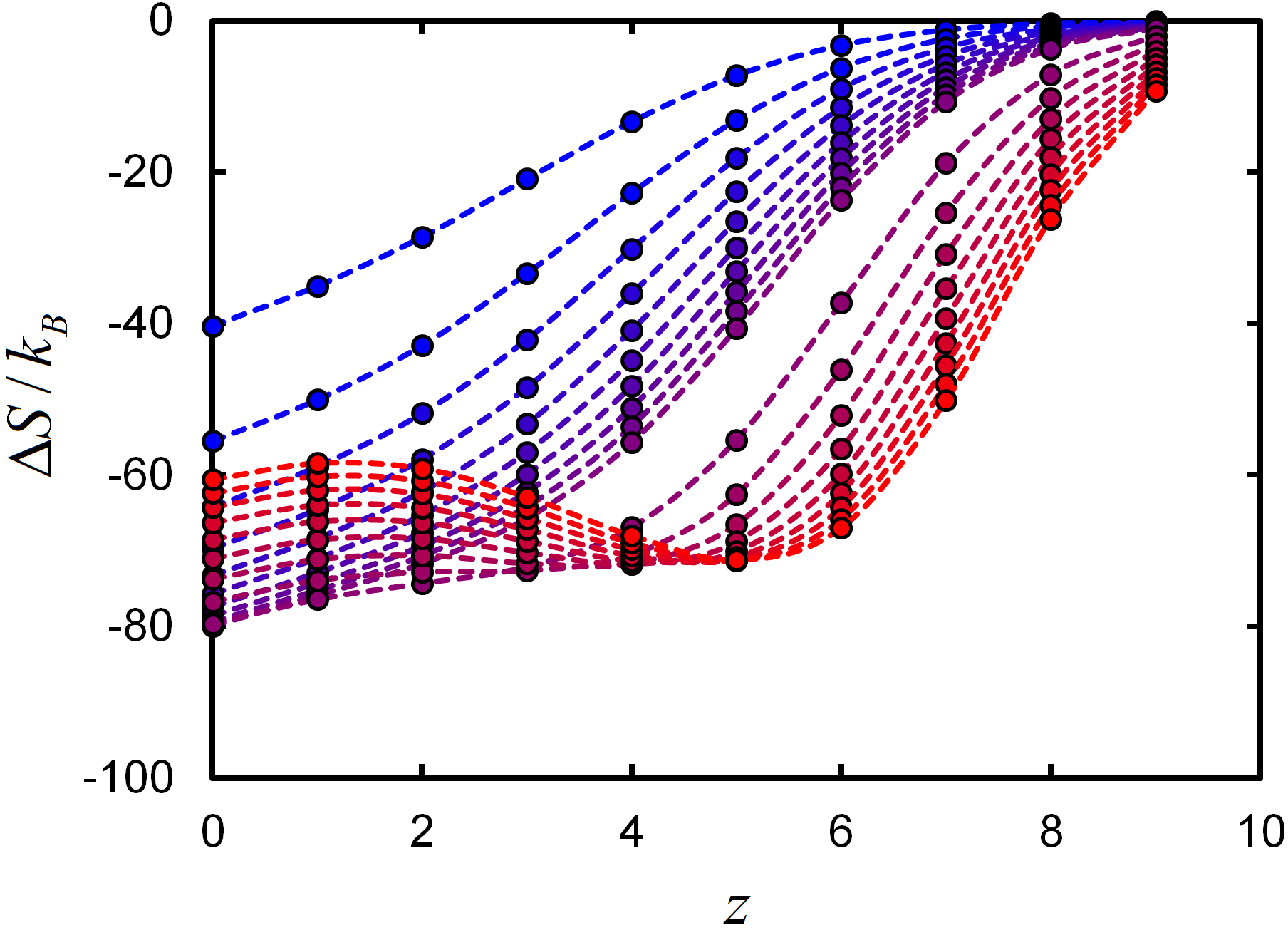}}
		\caption{Entropy of binding $\Delta S/k_B$ as a function of distance $z$ between receptor and particle surfaces, for different choices of receptor concentration $\phi_R$. See Figure \ref{fig:UvsH} for details.}
	\label{fig:SvsH}
\end{figure*}

\begin{figure*}
	\centering
		\subfigure[]{\includegraphics[width=0.495\textwidth]{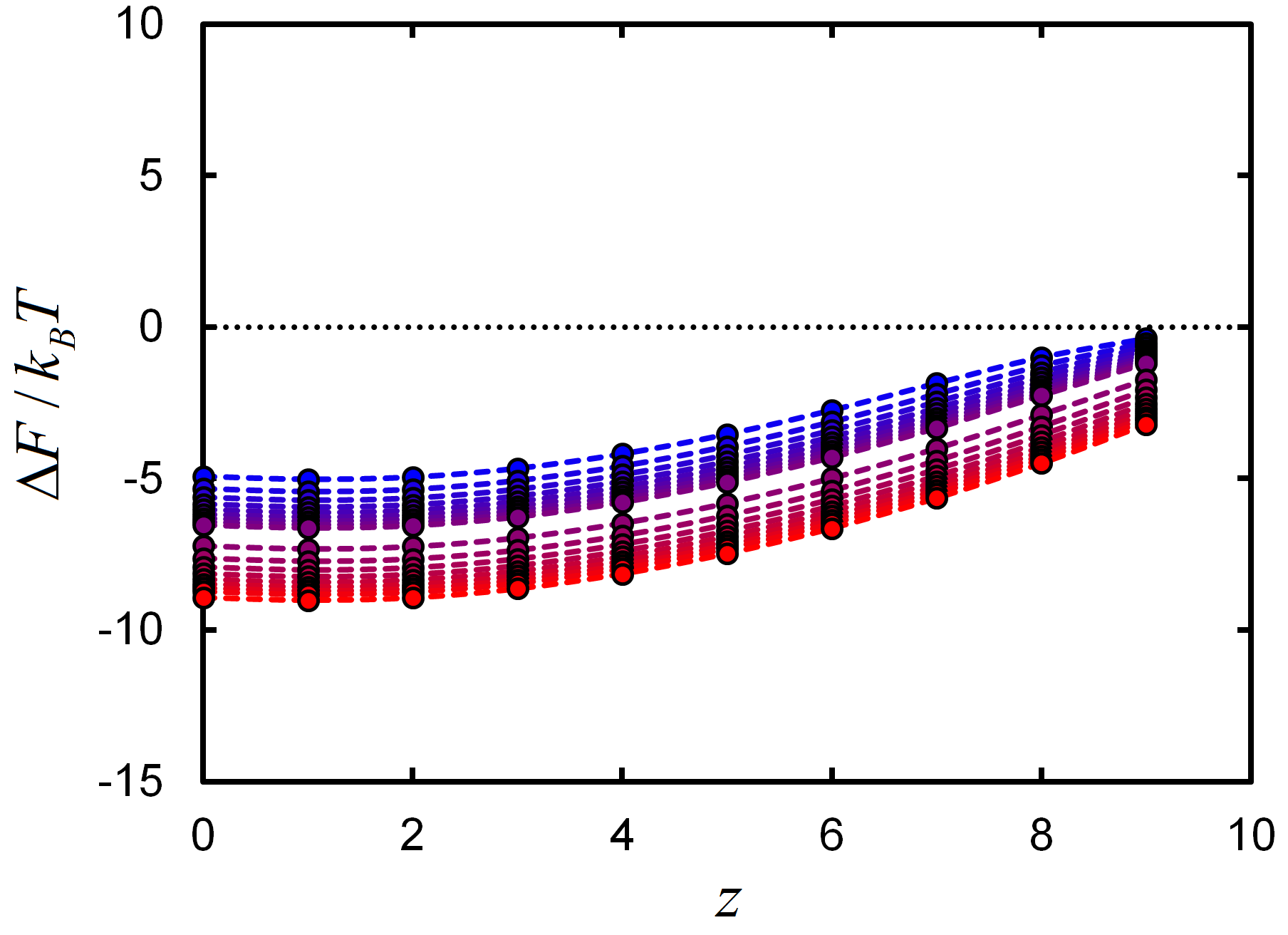}}
		\subfigure[]{\includegraphics[width=0.495\textwidth]{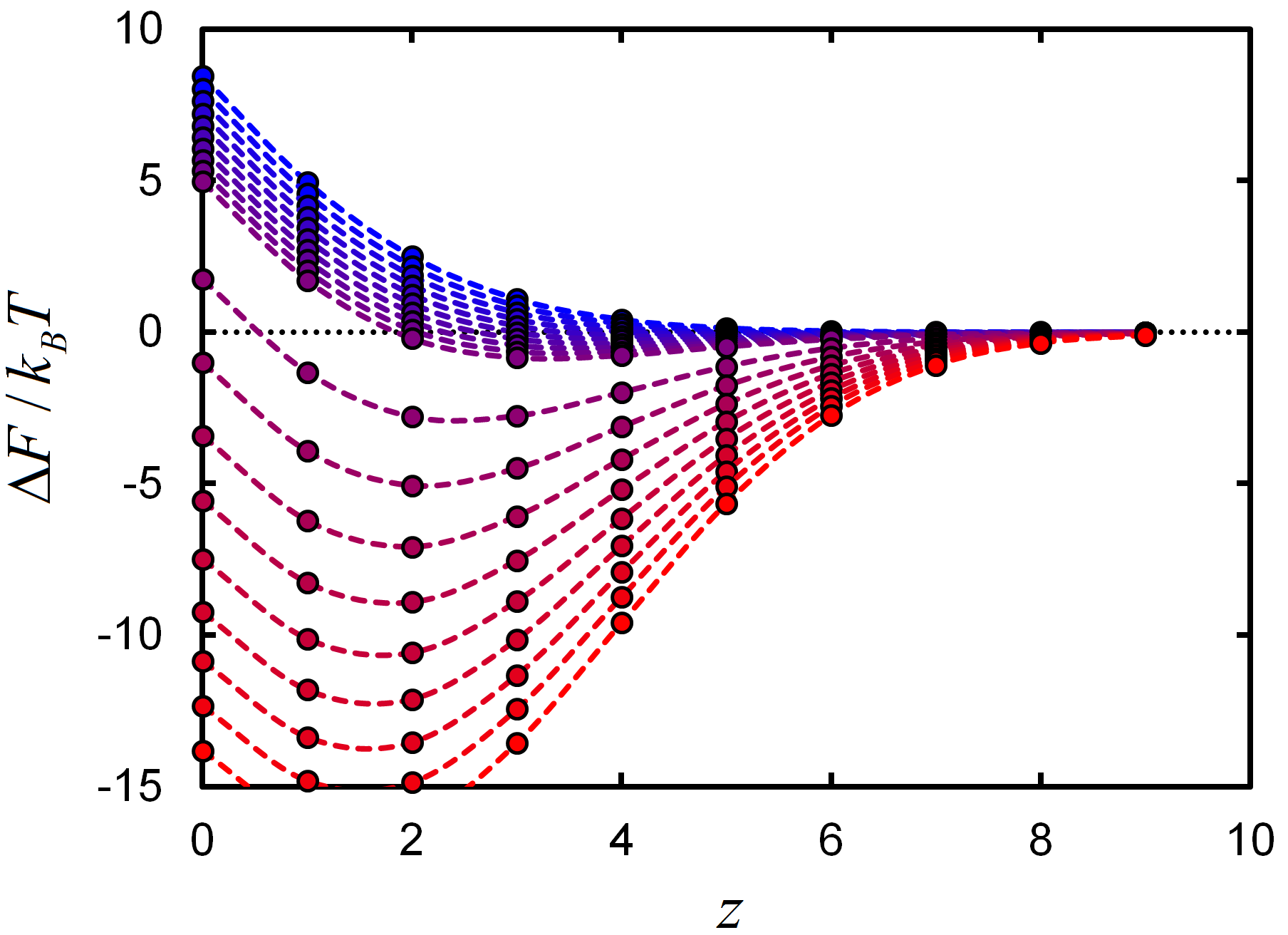}}
		\subfigure[]{\includegraphics[width=0.495\textwidth]{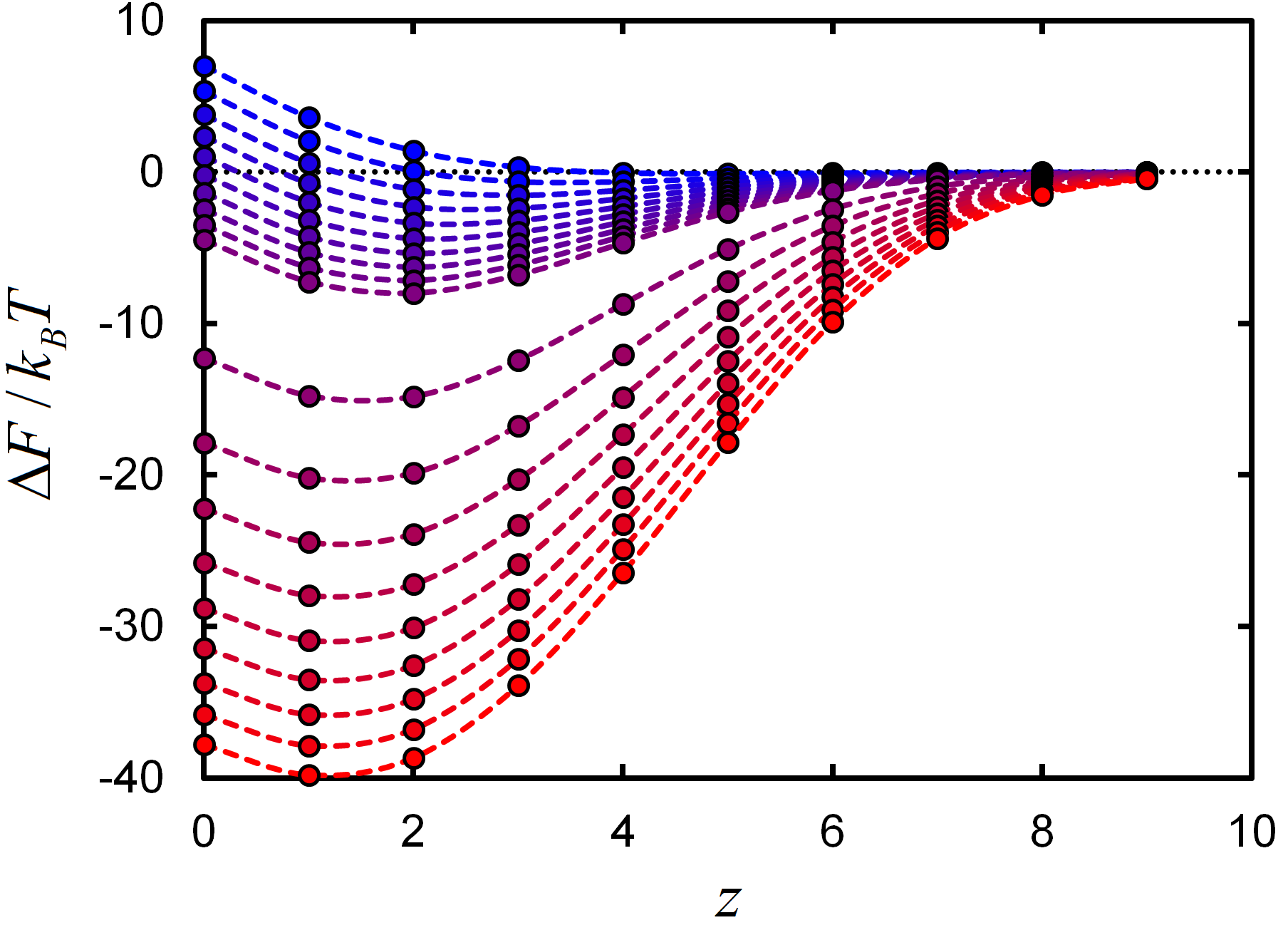}}
		\subfigure[]{\includegraphics[width=0.495\textwidth]{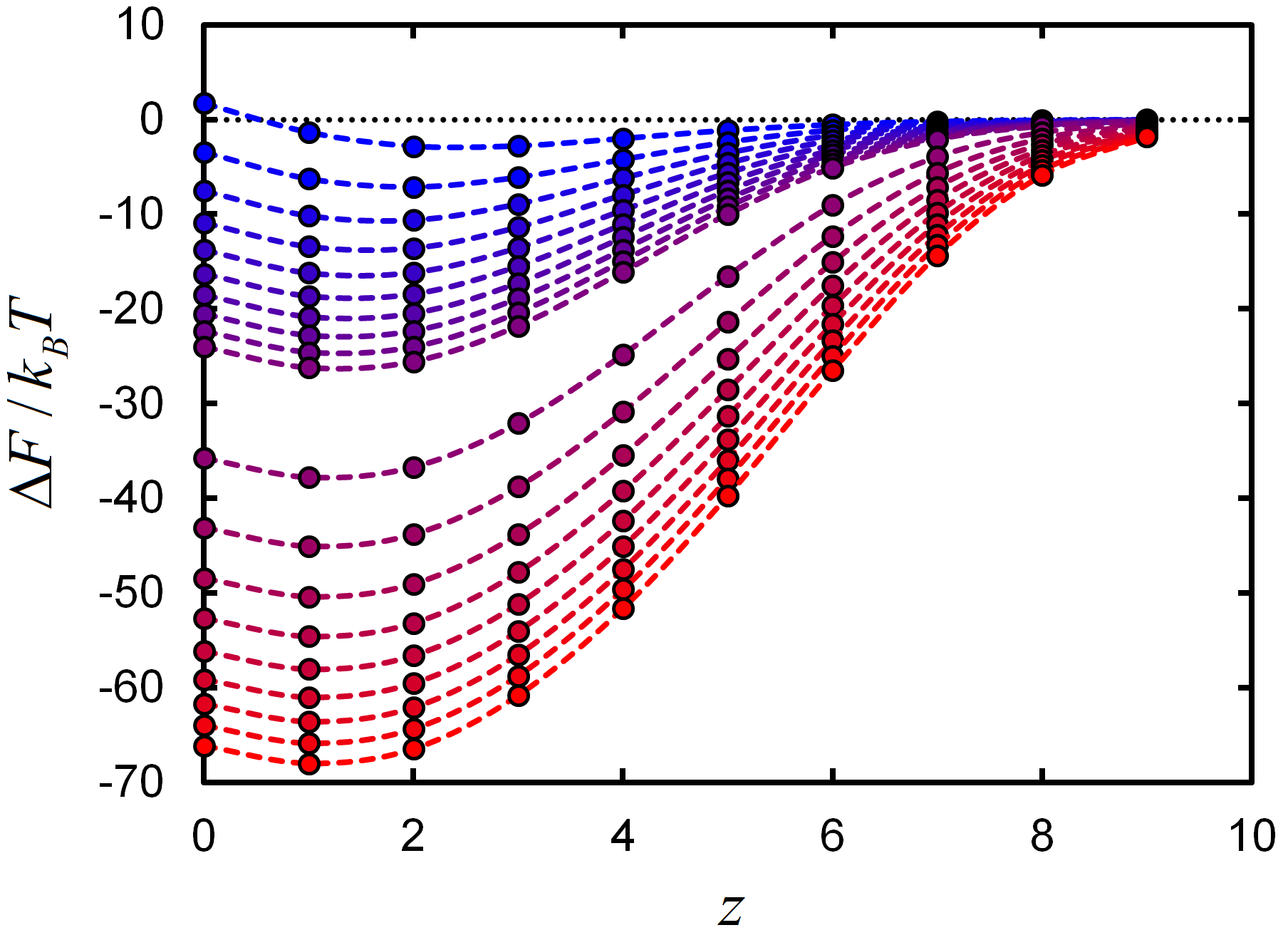}}
		\caption{Free energy of binding $\Delta F/k_B T$ as a function of distance $z$ between receptor and particle surfaces, for different choices of receptor concentration $\phi_R$. See Figure \ref{fig:UvsH} for details.}
	\label{fig:FEvsH}
\end{figure*}

\begin{figure*}
	\centering
		\subfigure[]{\includegraphics[width=0.495\textwidth]{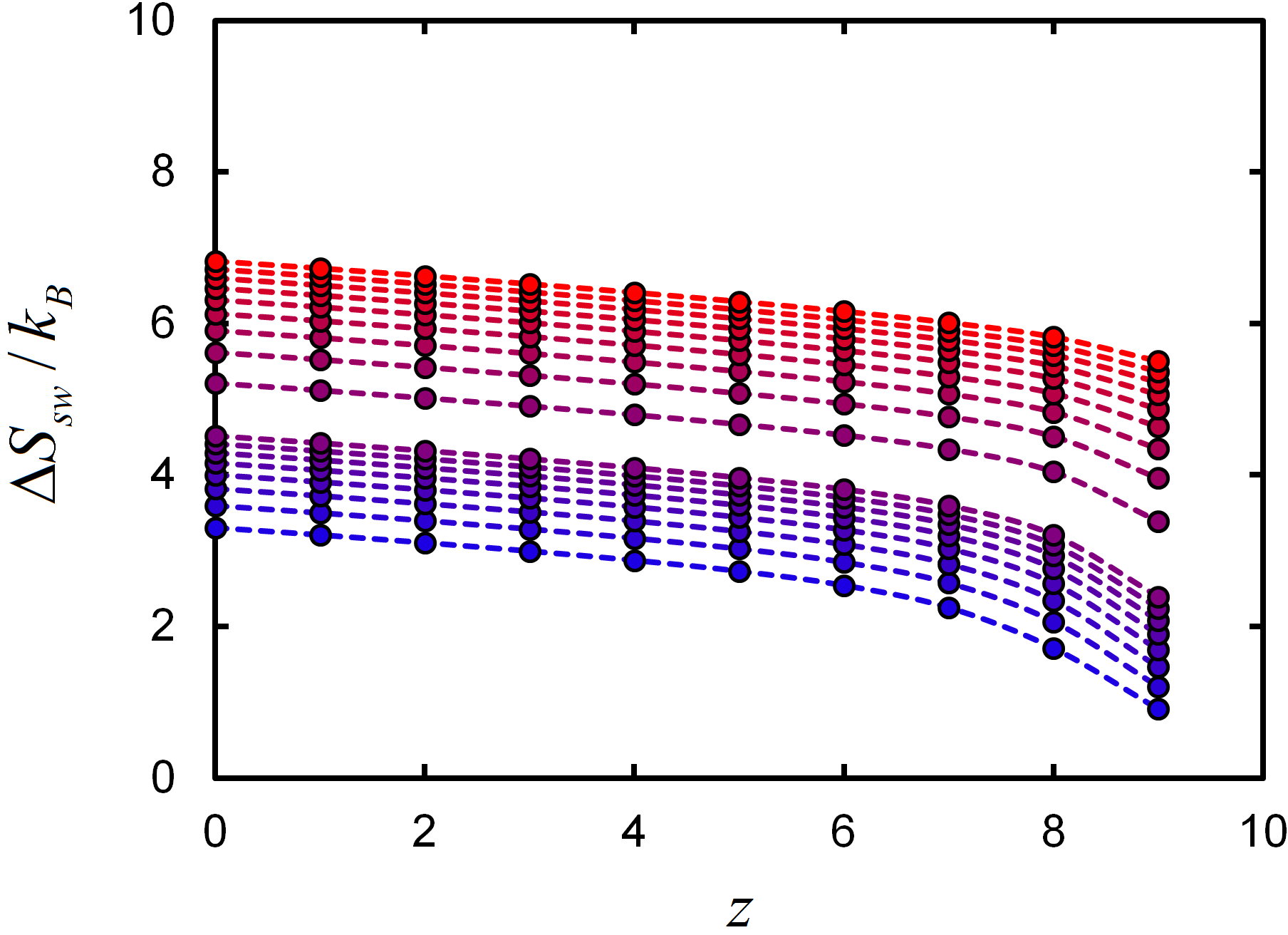}}
		\subfigure[]{\includegraphics[width=0.495\textwidth]{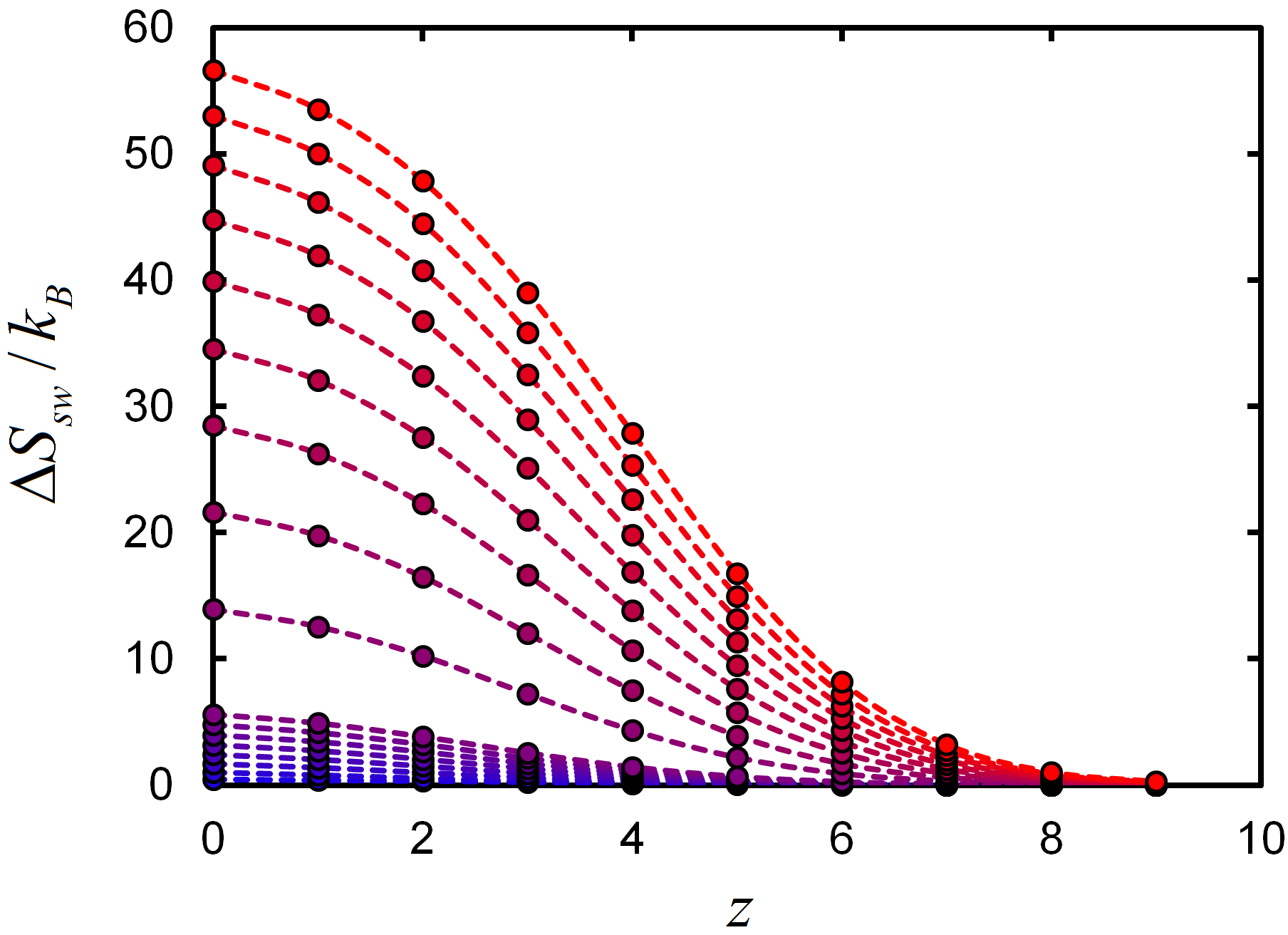}}
		\subfigure[]{\includegraphics[width=0.495\textwidth]{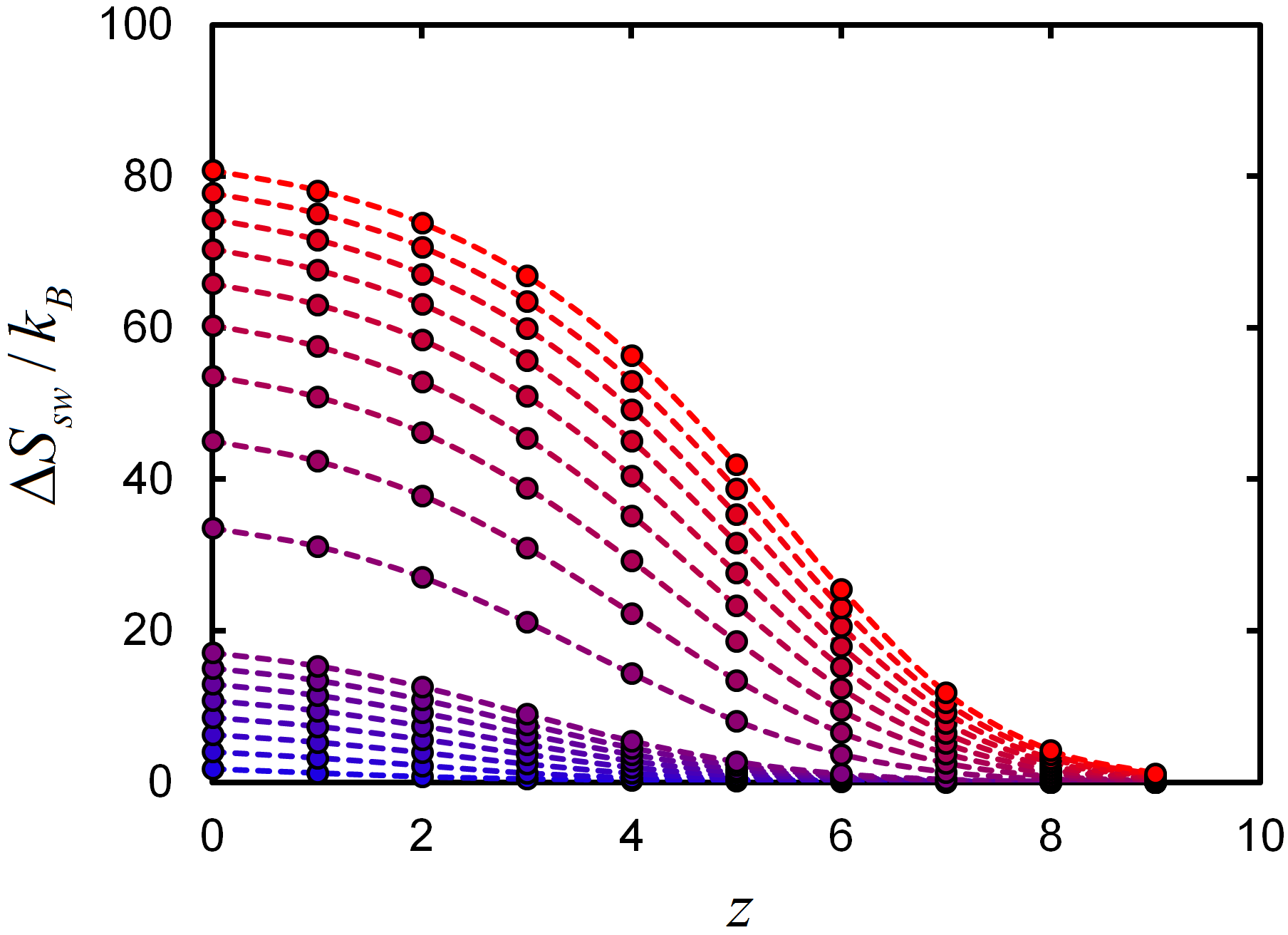}}
		\subfigure[]{\includegraphics[width=0.495\textwidth]{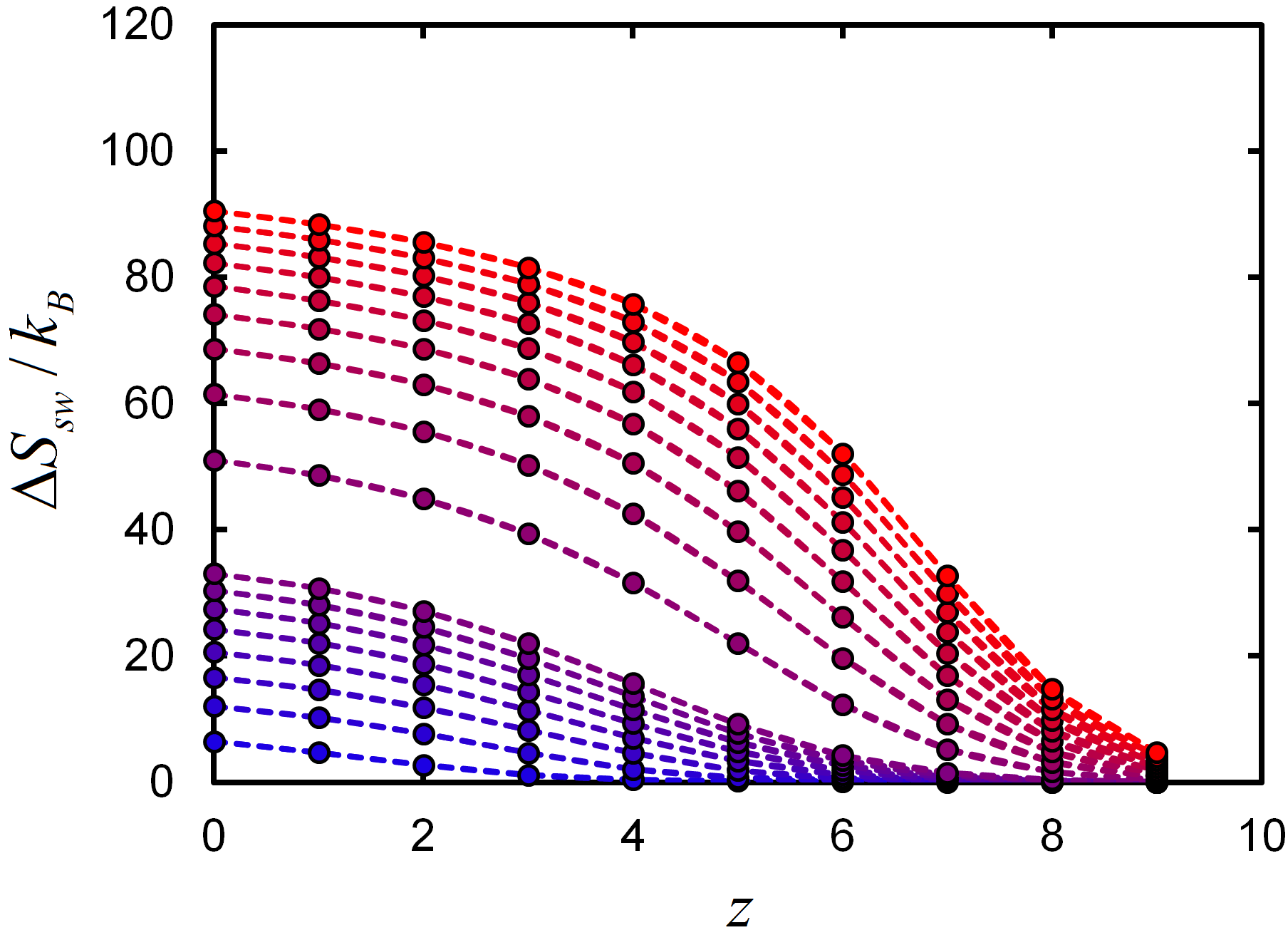}}
		\caption{Switchboard entropy $\Delta S_{sw}/k_B$ as a function of distance $z$ between receptor and particle surfaces, for different choices of receptor concentration $\phi_R$. See Figure \ref{fig:UvsH} for details.}
	\label{fig:SSWvsH}
\end{figure*}

For the $N_L = 1$ particle, the binding energy is very close to $\beta \epsilon = -12$ across $z$. This is the binding energy of the single ligand, indicating that the particle is in a bound state across the majority of $z$. Only when the receptor concentration is low, and the particle is located far from the surface, does the energy of binding become slightly less negative.

A selection of three different ligand-receptor binding energies $\beta \epsilon$ are examined for the $N_L = 20$ particle, in order to illustrate the sensitivity of the binding energy to this parameter. When $\beta \epsilon$ is relatively small ($-3.5$, in Figure \ref{fig:UvsH}b), the binding energy is strongly dependent on both $z$ and the receptor concentration. When the particle is nearest to the surface, at $z = 0$, the binding energy is maximised, as this is when the ligands are able to access the greatest number of receptors. By changing $z$, the number of available receptors decreases, reducing the number of bound ligands. 

When the surface is saturated with receptors, the minimum binding energy is near $-50 k_B T$; as the ligand-receptor binding energy is $-3.5 k_B T$ in this case, then there are $\approx 14$ bound ligands. The receptor concentration cannot be increased further, and so this represents the maximum number of bound ligands achievable for this choice of ligand-receptor binding energy. (The binding energy would be equal to $-70 k_B T$ were all ligands to be bound.)

Increasing $\beta \epsilon$, in Figures \ref{fig:UvsH}c and d, increases the number of bound ligands at smaller values of receptor concentration. Bringing the ligand-receptor binding energy to $-6.5$ (Figure \ref{fig:UvsH}d) leads the binding energy curves to asymptotically approach $\beta \Delta U = -130$---the case where all ligands are bound in that system. The curves also become flatter near small $z$, indicating that the number of bound ligands does not depend as strongly on the number of accessible receptors.

\begin{figure}
	\centering
		\subfigure[]{\includegraphics[width=0.495\textwidth]{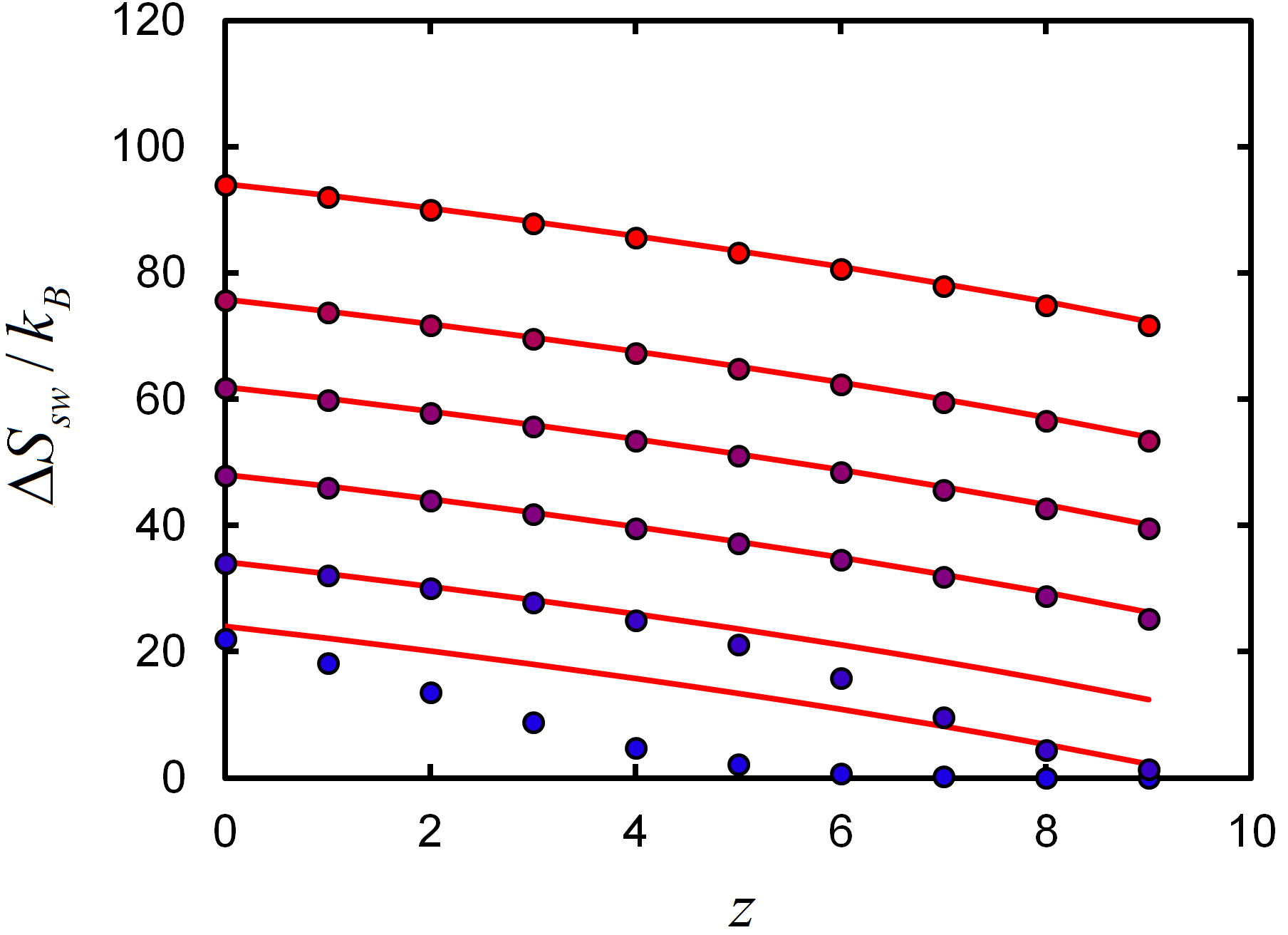}}
		\caption{Switchboard entropy $\Delta S_{sw}/k_B$ as a function of distance $z$ between receptor and particle surfaces in the limit that $\min{(N_R, N_L)}$ ligands are bound, for different choices of receptor concentration $\phi_R$. Points are calculations at each $z$ for $\phi_R = 0.03, 0.05, 0.1, 0.2, 0.4, 1.0$ (blue to red). In all cases $N_L = 20$, $N_{poly} = 20$, and particle $r = 2$. Solid red lines are plots of Eq. \ref{eqn:SSWPrediction} for each $\phi_R$.}
	\label{fig:SSWSBvsH}
\end{figure}

\begin{figure*}
	\centering
		\subfigure[]{\includegraphics[width=0.495\textwidth]{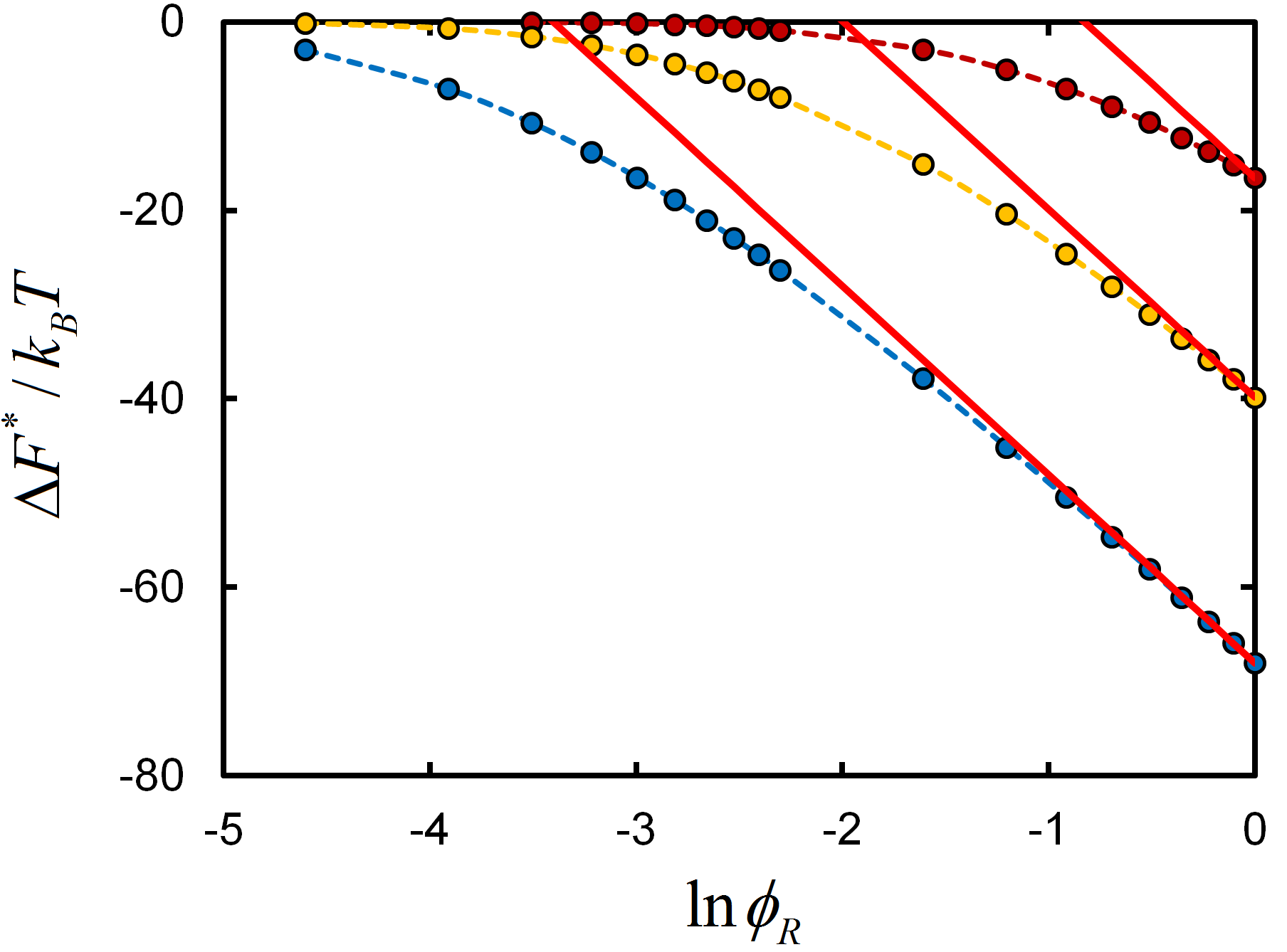}}
		\subfigure[]{\includegraphics[width=0.495\textwidth]{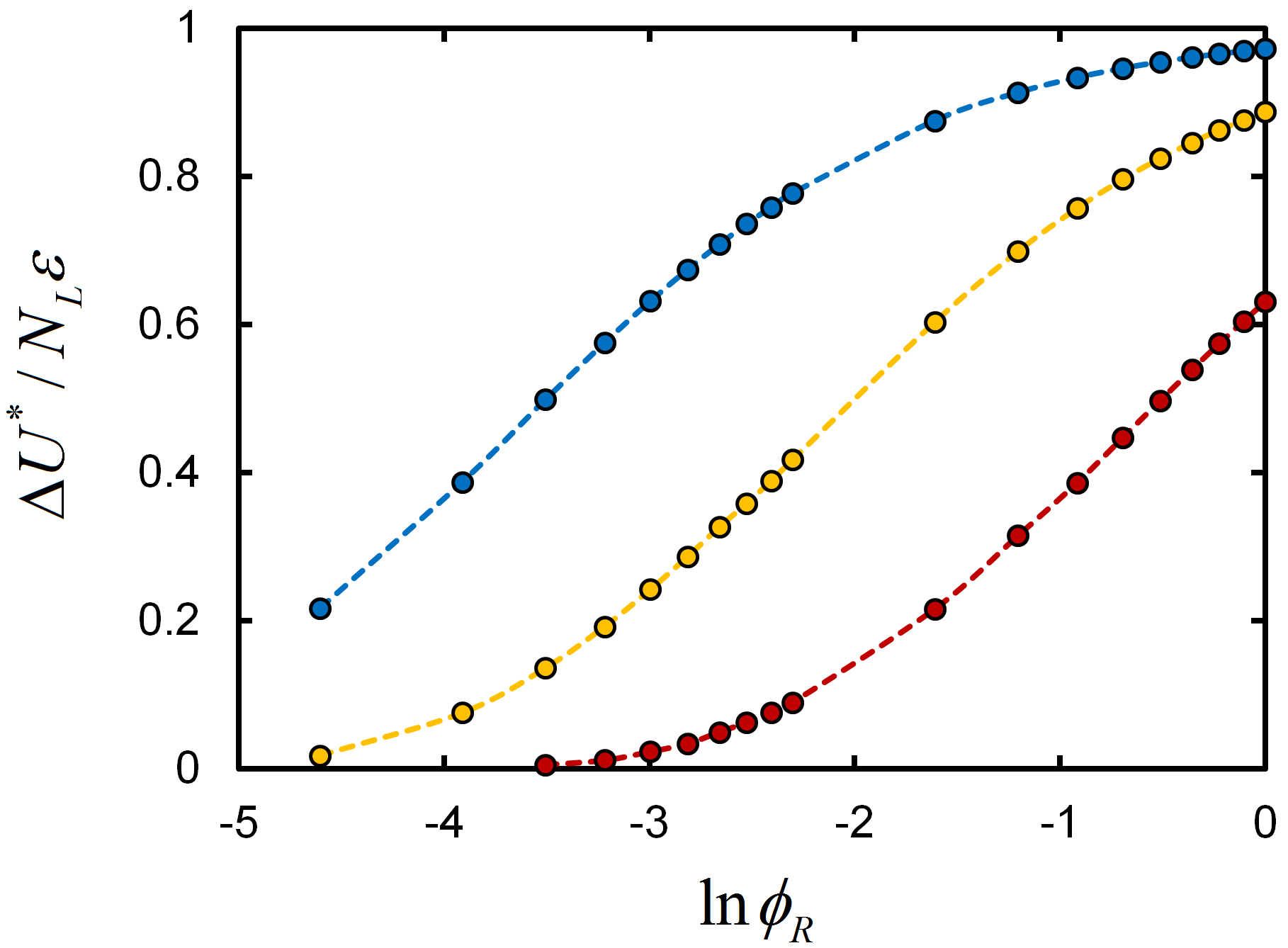}}
		\subfigure[]{\includegraphics[width=0.495\textwidth]{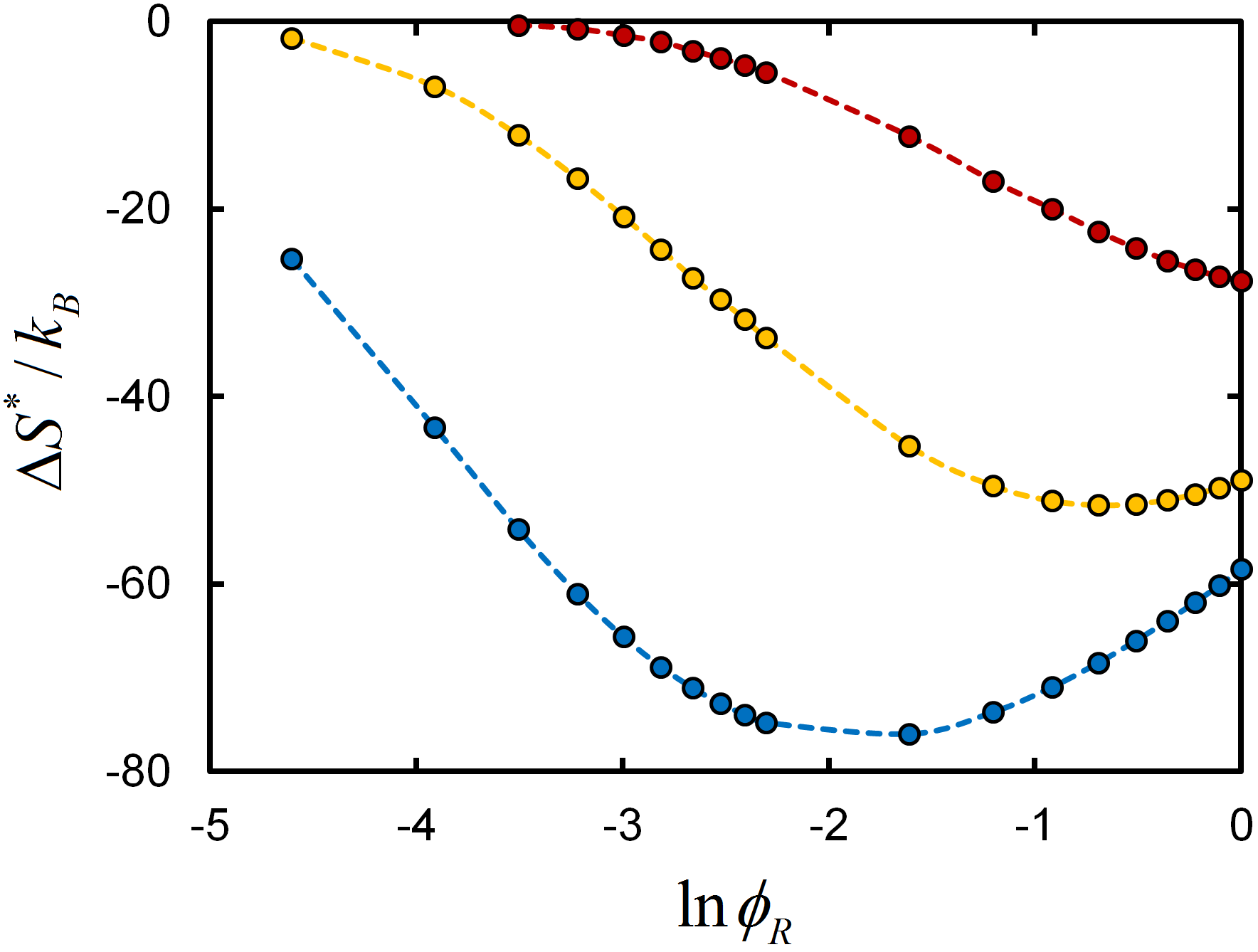}}
		\subfigure[]{\includegraphics[width=0.495\textwidth]{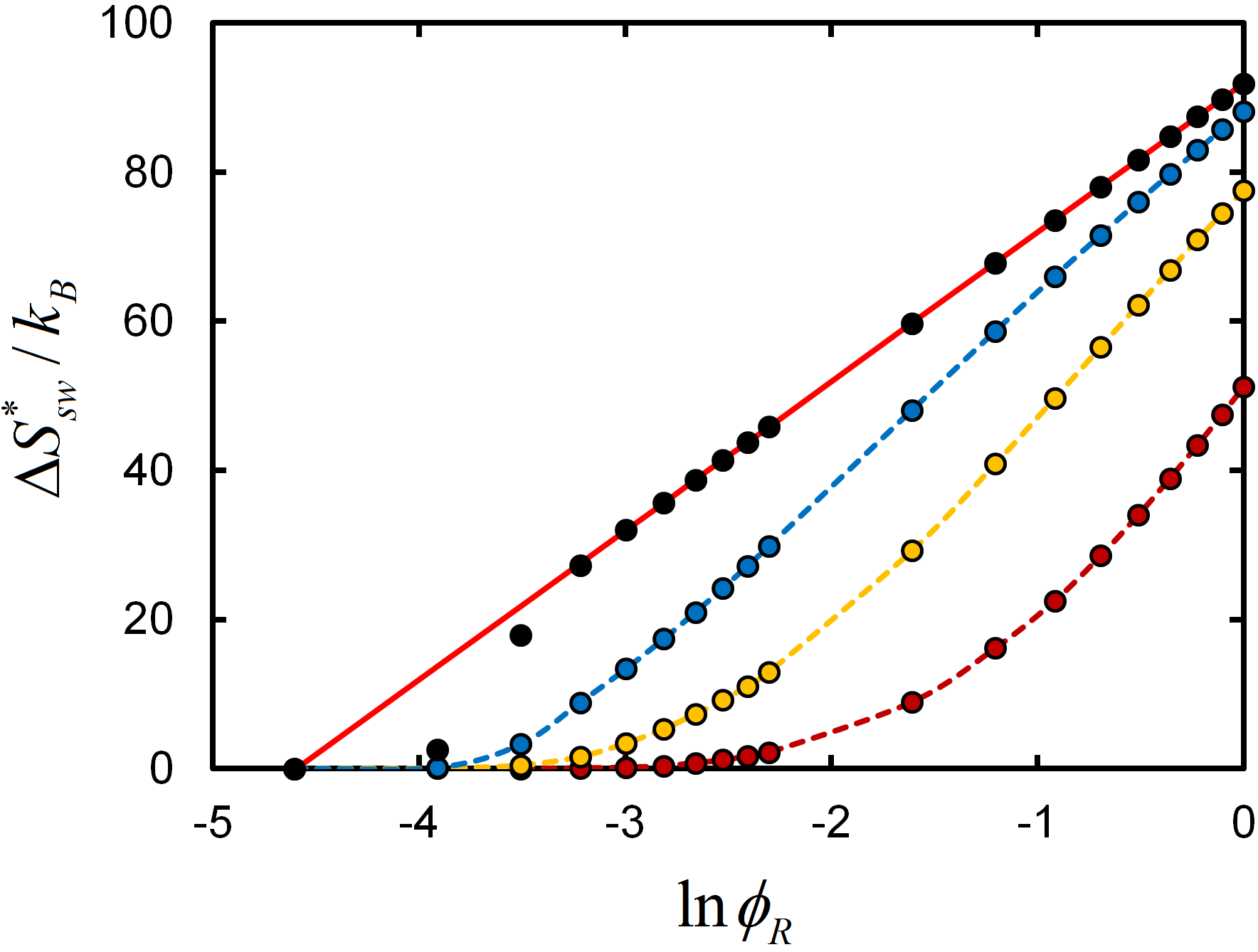}}
		\caption{Plots of the free energy $\Delta F^*$ (a), normalised energy $\Delta U^* / N_L \epsilon$ (b), entropy $\Delta S^*$ (c), and switchboard entropy $\Delta S_{sw}^*$ (d), at the equilibrium particle position $z^*$ as a function of the log of receptor concentration $\ln{\phi_R}$. Points are calculations, and dashed lines are fits to guide the eye. Parameters are $N_L = 20$, $N_{poly} = 20$, and $r = 2$ in all cases. Dataset in dark red has $\beta \epsilon = -3.5$, yellow has $\beta \epsilon = -5$, and blue has $\beta \epsilon = -6.5$. Data plotted in black in (d) are calculations in the strong-binding limit, when $\beta \epsilon \rightarrow \infty$. Solid red lines in (a) have slope of $-N_L$, and that in (d) has slope of $N_L$; vertical intercepts of lines have been shifted to match with numerical data.}
	\label{fig:MinimaVsLnFR}
\end{figure*}

Figure \ref{fig:SvsH} shows the binding entropy $\Delta S/k_B$ for each system in Figure \ref{fig:UvsH}. The binding entropy is calculated by
\begin{equation}
	\Delta S / k_B = -\beta \left(\Delta F - \Delta U\right),
	\label{eqn:BindingEntropy}
\end{equation}
where $\Delta F$ is given by Eq. 5. For the $N_L = 1$ particle, we observed in Figure \ref{fig:UvsH}a that the particle is predominantly in the bound state for low and high receptor concentrations. As a result, we find in Figure \ref{fig:SvsH}a that the entropy is similar for all choices of receptor concentration. Indeed, as the particle's single ligand is strongly-bound on average, then the entropy curves correspond to stretching or compressing the chain depending on the position $z$ of the particle core. For increasing receptor concentration, the entropy curve shifts slightly more positive, due to the larger number of binding sites for the ligand.  

The $N_L = 20$ systems in Figures \ref{fig:SvsH}b-d show strong dependence of the entropy on ligand-receptor binding energy. At low $\beta \epsilon$, the entropy becomes more negative with decreasing $z$. This is because the average number of bound ligands increases when the particle is closer to the surface; each ligand incurs an entropy penalty when it goes from an unbound to bound state. This is opposite from the behaviour shown in Figure \ref{fig:UvsH}b-d, where the binding energy is maximised when the particle is closest to the surface. The overall binding free energies $\Delta F$ in Figure \ref{fig:FEvsH} are a balance of these two factors, and the resulting $z$ that minimises the free energy is a nonzero value depending on the choice of $\phi_R$ and $\beta \epsilon$. 

Increasing $\beta \epsilon$ results in the formation of a local maximum in entropy at small $z$ in Figures \ref{fig:SvsH}c-d, becoming more pronounced at lower $\phi_R$ for larger $\beta \epsilon$. Comparing this behaviour with Figure \ref{fig:UvsH}b-d, we see that the formation of the maximum occurs when the rate of change of the total binding energy with $\phi_R$ slows down. Physically, this represents a turning point where the amount of entropy expended to increase the average number of bound ligands is less than that gained by the presence of additional receptors. As a result, entropy is dominated upon increasing $\phi_R$ by the addition of more receptors, through providing the bound ligands with more binding partners.

The presence of extra receptors is a type of degeneracy, providing the particle more binding possibilities without the expense of restricting additional ligands to the surface. We refer to this contribution as ``switchboard entropy''. Formally, we define the switchboard entropy as the extra amount of entropy arising from degenerate binding sites, compared to a case where each ligand may only bind to one receptor partner. Section \ref{sec:SwitchboardEntropy} discusses the method for calculating the switchboard entropy.

Figure \ref{fig:SSWvsH} plots the switchboard entropy for all cases examined in Figure \ref{fig:UvsH}. The quantity is always positive, and grows significantly as more receptors are added to the system. Comparing Figure \ref{fig:SSWvsH}a to Figures \ref{fig:SSWvsH}b-d, we see that the multivalent $N_L = 20$ particle has a substantially larger switchboard entropy contribution, compared to the monovalent $N_L = 1$ particle.

When the system approaches the limit that all ligands are strongly bound to a large number of receptors, then the entropy of binding should follow the theoretical prediction derived in Eq. 1:
\begin{equation}
	\Delta S_{sw}(z)/k_B \approx N_L \ln{\bar{N}_R(z)} - \ln{N_L!}.
	\label{eqn:SSWPrediction}
\end{equation}
(Note that the extra factor of $N_L!$ accounts for indistinguishability of the ligands.) Figure \ref{fig:SSWSBvsH} shows examples of the switchboard entropy calculated for systems that have strongly-bound ligands, showing excellent agreement with Eq. \ref{eqn:SSWPrediction} in this limit.

\section{Behaviour at the free energy minimum}

We now turn to the behaviour of the binding thermodynamics at the equilibrium particle position, depending on the mean receptor concentration $\phi_R$. Figure \ref{fig:MinimaVsLnFR} shows plots of the free energy, entropy, switchboard entropy, and energy of the multivalent particle at its equilibrium position $z^*$ for various $\ln{\phi_R}$. Results are shown for particles with $N_L = 20$ ligands, and ligand-receptor binding energies of $\beta \epsilon = -3.5, -5$, and $-6.5$. 

Considering plots of the free energy in Figure \ref{fig:MinimaVsLnFR}a, as the ligand-receptor binding strength increases the portion of the curve that is linear grows larger. The red lines shown in Figure \ref{fig:MinimaVsLnFR}a have slopes $N_L$; evidently, as $\beta \epsilon$ grows more negative, the free energy at high receptor concentration goes as
\begin{equation}
	\beta \Delta F^* \propto -N_L \ln{\phi_R}.
	\label{eqn:FEScaling}
\end{equation}
This is the result anticipated by the simple Langmuir theory, Eq. 1. The transition into this regime occurs when the total energy of the particle approaches its saturation value, $N_L \epsilon$. We can examine this by plotting the scaled total particle energy $\Delta U^* / N_L \epsilon$, shown in Figure \ref{fig:MinimaVsLnFR}b. As $\beta \epsilon$ and the receptor concentration grow large, the total binding energy approaches its maximum value.

Because the total binding energy in Figure \ref{fig:MinimaVsLnFR}b approaches a constant for increasing $\beta \epsilon$ and $\phi_R$, then it must be the entropy of binding that gives rise to the change in binding free energy following Eq. \ref{eqn:FEScaling}. 

Accordingly, Figure \ref{fig:MinimaVsLnFR}c gives plots of the binding entropy at equilibrium, as a function of $\ln{\phi_R}$. The striking non-monotomic behaviour of these curves reflects the conclusions we drew from Figure \ref{fig:SvsH}. 

When the receptor concentration is low, the ligands of the particle are in equilibrium with few binding partners. For moderate $\beta \epsilon$, the translational entropy of an unbound ligand is quite large compared to the energy it gains upon binding. Therefore, not all ligands are bound. 

Upon adding more receptors in this regime, the ligand equilibrium is further biased towards receptor binding; however, the binding of each ligand costs the particle additional entropy. Eventually, at a sufficient receptor concentration, all ligands become bound. We then enter a regime in which adding additional receptors \emph{increases} the binding entropy, by providing degenerate binding partners for the bound ligands. This is the switchboard entropy. Since all ligands are already bound, the particle does not spend any additional entropy to restrict ligands to the surface. 

This behaviour, specific to a multivalent (as opposed to monovalent) interaction, gives rise to the unique non-monotonic entropy of binding as shown in Figure \ref{fig:MinimaVsLnFR}c. A monovalent particle exhibits similar behaviour at low receptor concentration, as the ligand state equilibrium is balanced in the same way between translational entropy and receptor binding. However, once the equilibrium is sufficiently biased towards the bound state, there is little effect obtained by adding degenerate binding partners. (We saw this in Figure \ref{fig:SSWSBvsH}a, where there is only a very small switchboard entropy contribution to the binding of a particle with just one ligand.)

Figure \ref{fig:MinimaVsLnFR}d plots the switchboard contribution to the binding entropy, as a function of receptor concentration. Upon increasing $\phi_R$ and $\beta \epsilon$ to the strong-binding limit (going from red to black datasets), the behaviour of the switchboard entropy approaches 
\begin{equation}
	\Delta S_{sw}^*/k_B \propto N_L \ln{\phi_R},
	\nonumber
\end{equation}
as found in Figure \ref{fig:SSWvsH} and Eq. \ref{eqn:SSWPrediction}. The switchboard entropy, then, is the primary factor driving the behaviour of the free energy at large $\beta \epsilon$ and $\phi_R$ in Figure \ref{fig:MinimaVsLnFR}a.

\section{Calculating ``switchboard'' entropy}
\label{sec:SwitchboardEntropy}

To calculate the entropy arising from the presence of extra receptors on the surface (in the ``real'' system), we must calculate the entropy for an equivalent system where each of the $N_L$ ligands only have one possible receptor partner. We will refer to this as the ``reference'' system.

The reference system is different depending on the number of receptors $N_R$ in the real system. If the real system has $N_R \leq N_L$, then the reference system is the same as the real system. This is because the ligands are still considered indistinguishable in the reference system, leading to the fact that each ligand can still bind to any one of the available receptors. This is identical to the real system. 

On the other hand, if $N_R > N_L$ in the real system, then the reference system has exactly $N_R = N_L$ receptors in order to remove the excess receptors. The difference in the average entropy between these two systems represents the contribution arising from the presence of extra binding partners in the real system---the switchboard entropy.

We will start by showing the calculation for switchboard entropy given one receptor configuration, and then perform an average over all possible configurations given by the distribution $P(N_R)$ of the number of receptors afterwards.

Regardless of $N_R$ in the real system, the reference system must have the same average number of bound ligands as the real system. (Otherwise, the difference in entropy between the two systems would contain extra factors not due to excess receptors.) Thus, we must first determine the average number of bound ligands in the real system, and then adjust the parameters of the reference system accordingly.

The average number of bound ligands in one real system is calculated by
\begin{equation}
	N_{b,real} = \frac{U_{real}}{\epsilon_{real}},
	\nonumber
\end{equation}
where $\epsilon_{real}$ is the ligand-receptor binding energy in the real system. The entropy is obtained by
\begin{equation}
	\frac{S_{real}}{k_B} = -\beta \left(F_{real} - U_{real}\right).
	\nonumber
\end{equation}
For $N_R \leq N_L$ in the real system, then the reference system is identical to the real system. When $N_R > N_L$, then the reference system has $N_R = N_L$ receptors, and must exhibit the same average number $N_{b,real}$ of bound ligands as in the real system. The value of $\beta \epsilon$ in the reference system can be adjusted to achieve $N_{b,ref} = N_{b,real}$. Another method is to perform two calculations of the reference system, for when there are exactly $N_{b,ref} = \mbox{floor}(N_{b,real})$ and then $\mbox{ceiling}(N_{b,real})$ bound ligands. The entropy for the true reference system is then
\begin{align}
	S_{ref} = &\left(1 - p\right) S_{ref}(N_b = \mbox{floor}(N_{b,real})) \nonumber \\
	&+ p S_{ref}(N_b = \mbox{ceiling}(N_{b,real}))
	\label{eqn:SRef}
\end{align}
where
\begin{equation}
	p = N_{b,real} - \mbox{floor}(N_{b,real}).
	\nonumber
\end{equation}
In principle, if our real system has $N_R > N_L$ receptors, then many reference systems with $N_R = N_L$ could be constructed, each having a different arrangement of the $N_R$ receptors (and therefore a different $S_{ref}$). However, we do not need to be concerned about making a choice here, as we now perform an average over all possible receptor configurations.

The number of receptors on the surface follows the distribution $P(N_R)$; for a given $N_R$, the receptors are randomly distributed across the $N_A$ surface sites. The switchboard entropy averaged over all of these configurations can be written
\begin{equation}
	\bar{S}_{sw} = \frac{1}{N} \sum_{n}{\left(S_{real,n} - S_{ref,n}\right)},
	\label{eqn:SwEntropyTrialAvg}
\end{equation}
where trials are indexed by $n$. The quantity $S_{real,n}$ is the entropy of the real system on trial $n$. The reference entropy $S_{ref,n}$ on trial $n$ takes one of two values, depending on the value of $N_R$ in that trial:
\begin{equation}
	S_{ref,n} = 
	\begin{cases}
    S_{ref} ,& \text{if } N_R > N_L\\
    S_{real,n},              & \text{if } N_R \leq N_L
	\end{cases}
	\nonumber
\end{equation}
Thus, only trials having $N_R > N_L$ will contribute to the sum in Eq. \ref{eqn:SwEntropyTrialAvg}. If we define $N^+$ to be the number of trials with $N_R > N_L$, then the switchboard entropy can be calculated by absorbing the average over $n$ into the two entropy terms in Eq. \ref{eqn:SwEntropyTrialAvg}:
\begin{equation}
	\bar{S}_{sw} = \frac{N^+}{N} \left(\bar{S}_{real,+} - \bar{S}_{ref} \right).
	\nonumber
\end{equation}
This is the equation used to calculate the average switchboard entropies given in Figure \ref{fig:SSWvsH}. The expression is conveniently in terms of two averages over all receptor configurations: 
\begin{itemize}
	\item{$\bar{S}_{real,+}$ is the average entropy for (real) systems having $N_R > N_L$ receptors;}
	\item{$\bar{S}_{ref}$ is the average entropy for (reference) systems having exactly $N_R = N_L$ receptors and $\bar{N}_{b,real}$ bound ligands, where $\bar{N}_{b,real}$ is the number of bound ligands averaged over the real systems.}
\end{itemize}
The prefactor $N^+/N$ can be calculated analytically given $P(N_R)$:
\begin{equation}
	\frac{N^+}{N} = 1 - \sum_{N_R = 0}^{N_L}{P(N_R)}
	\nonumber
\end{equation}

To calculate the quantities $\bar{S}_{real,+}$ and $\bar{S}_{ref}$ in practise, we derive relevant forms of the partition function given by Eq. 3, calculate the free energy, and then obtain the entropy by Eq. \ref{eqn:BindingEntropy}. For $\bar{S}_{real,+}$, we compute $Q_{full}$ with $P(N_R)$ renormalised to reside between $N_R = N_L + 1$ and $N_A$. (Obviously, the particle must be positioned so that $N_A > N_L$.) If we define
\begin{equation}
	\tilde{Q}(\lambda) = \frac{\binom{N_L}{\lambda}}{\binom{N_A}{\lambda}} \lambda! Q_{ub}(\lambda) Q_b(N_A, \lambda)
	\nonumber
\end{equation}
then we can calculate the average free energy $\beta \bar{F}_{real,+}$ by
\begin{equation}
	\beta \bar{F}_{real,+} = -\ln{} \sum_{\lambda = 0}^{N_L}    \tilde{Q}(\lambda)    \left(\sum_{N_R = N_L + 1}^{N_A}{\binom{N_R}{\lambda} P'(N_R)}\right).
	\nonumber
\end{equation}
The renormalised distribution $P'(N_R)$ for $N_R$ is
\begin{equation}
	P'(N_R) = \frac{P(N_R)}{\sum_{N_R' = N_L + 1}^{N_A}{P(N_R')}}.
	\nonumber
\end{equation}
To compute $\bar{S}_{ref}$, we force the distribution $P(N_R)$ to collapse to a Kronecker delta function at $N_R = N_L$. The average free energy across all receptor configurations in this case takes the form
\begin{equation}
	\beta \bar{F}_{ref} = -\ln{} \sum_{\lambda = 0}^{N_L}   \tilde{Q}(\lambda)  \binom{N_L}{\lambda}.
	\nonumber
\end{equation}
We then compute $\bar{S}_{ref}$ using Eq. \ref{eqn:SRef}, which requires evaluating $\bar{F}_{ref}$ for $N_b = \mbox{floor}(N_{b,real})$ and $\mbox{ceiling}(N_{b,real})$ bound ligands, respectively. The equation for $\bar{F}_{ref}$ for a set number $N_b$ of bound ligands is
\begin{equation}
	\beta \bar{F}_{ref}(N_b) = -\ln{} \left[ \tilde{Q}(N_b)  \binom{N_L}{N_b} \right].
	\nonumber
\end{equation}
The switchboard entropy is then obtained by Eq. \ref{eqn:BindingEntropy} .

\end{document}